\documentclass[11pt,journal,draftcls,onecolumn]{IEEEtran}
\IEEEoverridecommandlockouts
\usepackage{url}
\usepackage{epsf}
\usepackage{graphicx}
\usepackage{ifthen}
\usepackage{amsmath,amssymb,amsfonts}
\usepackage{color}
\usepackage{subfigure}
\usepackage{fancyhdr}
\usepackage{verbatim,moreverb}
\usepackage{pifont}
\usepackage{setspace,supertabular}
\usepackage{mathrsfs}
\usepackage{psfrag}
\usepackage{bbm}

\DeclareMathOperator*{\mycup}{\cup}

\newcommand{\mybinom}[3]{\genfrac{(}{)}{0pt}{#3}{#1}{#2}}
\newtheorem{thm}{Theorem}
\newtheorem{lem}[thm]{Lemma}
\newtheorem{prop}[thm]{Proposition}

\newtheorem{cor}[thm]{Corollary}
\newtheorem{defn}{Definition}[section]

\newcommand{\bb}[1]{\mathbf{#1}}
\def\Power{{\mathrm P}}

\newcommand{\C}{\mathcal{C}}
\newcommand{\Reals}{\mathbb{R}}

\newcommand{\Ca}[1]{\mathscr{C}_{#1}}
\newcommand{\Ro}[1]{\overline{\mathscr{C}}_{#1}}
\newcommand{\Ri}[1]{\underline{\mathscr{C}}_{#1}}

\newcommand{\Scal}[4]{ \mathcal{S}}

\newcommand{\R}[3]{%
\ifthenelse{\equal{#1}{}}{R^{#3}(#2)}{R^{#3}_{#1}\left(#2 \right)}%
} 
\newcommand{\rh}[1]{ \rho_{#1}  }
\newcommand{\rhb}[1]{ \pmb{\rho}_{#1} }
\newcommand{\W}[3]{%
\ifthenelse{\equal{#1}{}}{W^{#3}(#2)}{W^{#3}_{#1}\left(#2 \right)}%
} 
\newcommand{\hatW}[3]{%
\ifthenelse{\equal{#1}{}}{\hat{W}^{#3}(#2)}{\hat{W}^{#3}_{#1}\left(#2 \right)}%
}
\newcommand{\X}[2]{%
\ifthenelse{\equal{#2}{n}}{\bb{X}_{#1}}{X_{#1,#2}}%
} 
\newcommand{\Y}[2]{%
\ifthenelse{\equal{#2}{n}}{\bb{Y}_{#1}}{Y_{#1,#2}}%
} 
\newcommand{\D}[2]{\mathcal{W}_{#1}(#2)}
\newcommand{\Ra}[2]{ r_{#1}(#2)}

\newcommand{\plo}[3]{p_{#2}(#3)}
\newcommand{\pup}[3]{ p_{#2}}

\newcommand{\wwp}[1]{W_{#1,1}}
\newcommand{\wwc}[1]{W_{#1,2}}
\newcommand{\hwp}[1]{ \hat{W}_{#1,1}}
\newcommand{\hwc}[1]{ \hat{W}_{#1,2}}

\newcommand{\Rp}[1]{R_{#1,1}}
\newcommand{\Rc}[1]{R_{#1,2}}

\begin{document}

\title{Random Access: An Information-Theoretic Perspective} 

\author{Paolo~Minero, Massimo~Franceschetti, and David~N.~C. Tse
\thanks{P. Minero and M. Franceschetti are with the Advanced Network Science group (ANS) of the California Institute of Telecommunications and Information Technologies (CALIT2), Department of Electrical and Computer Engineering, University of California, San Diego CA, 92093, USA. Email: pminero@ucsd.edu, massimo@ece.ucsd.edu.}%
\thanks{David~N.~C.~Tse is with the Wireless Foundations at the Department of Electrical Engineering and Computer Science, University of California, Berkeley, CA 94720, USA.
Email: dtse@eecs.berkeley.edu.
}
\thanks{This work was partially supported by the National Science Foundation CAREER award CNS-0546235, and CCF-0635048, and by the Office of
Naval Research YIP award N00014-07-1-0870.}
}
\maketitle

\begin{abstract}
This paper considers a random access system where each sender can be in two modes of operation, active or not active, and where the set of active users is available to a common receiver only. Active transmitters encode data into independent streams of information, a subset of which are decoded by the receiver, depending on the value of the collective interference. The main contribution is to present
an information-theoretic formulation of the problem which allows us to characterize, with a guaranteed gap to optimality, the rates that can be achieved by different data streams.

Our results are articulated as follows. First, we exactly characterize the capacity region of a two-user system assuming a binary-expansion deterministic channel model. Second, we extend this result to a two-user additive white Gaussian noise channel, providing an approximate characterization within $\sqrt{3}/2$ bit of the actual capacity. Third, we focus on the \emph{symmetric} scenario in which users are active with the same probability and subject to the same received power constraint, and study the maximum achievable expected sum-rate, or throughput, for \textit{any} number of users. In this case,  for the symmetric binary expansion deterministic channel (which is related to the packet collision model used in the networking literature), we  show that a simple coding scheme which does \emph{not} employ superposition coding achieves the system throughput.  This result also  shows that  the performance of slotted ALOHA systems can be improved by allowing encoding rate adaptation at the transmitters, achieving constant  (rather than zero) throughput as the number of users tends to infinity. For the symmetric additive white Gaussian noise channel,  we propose a scheme that is within one bit of  the system throughput for any value of the underlying parameters.
\end{abstract}

\newpage

\section{Introduction}

Random access is one of the most commonly used medium access control schemes for channel sharing by a number of transmitters.
Despite decades of active research in the field, the theory of random access communication is far from complete. What has been   notably pointed out by Gallager in his review paper  more than two decades ago~\cite{Gallager} is still largely true:
on the one hand, information theory provides accurate models for the noise and for the interference caused by simultaneous transmissions, but it ignores random message arrivals at the transmitters; on the other hand, network oriented studies focus on the  bursty nature of messages, but do not accurately describe the underlying physical channel model. As an example of the first approach,
the classic results by Ahlswede~\cite{Ahlswede0} and Liao~\cite{Liao} provide a complete characterization of the set of rates that can be simultaneously achieved communicating over a discrete memoryless (DM) multiple access channel (MAC). But the coding scheme they develop assumes a fixed number of transmitters with continuous presence of data to send.  As an example of the second approach, Abramson's classic collision model for the ALOHA network~\cite{Abramson} assumes that packets are transmitted at random times and are encoded at a \textit{fixed} rate, such that a packet collision occurs whenever two or more transmitters are simultaneously active. The gap between these two lines of research is notorious and well documented by Ephremides and Hajek in their survey article \cite{Ephremides}.

In this paper, we try to bridge the divide between the two  approaches described above. We present the analysis of a model which is information-theoretic in nature, but which also accounts for the random activity of users, as in models arising in the networking literature. We consider a crucial aspect of random access, namely that the number of simultaneously transmitting users is unknown to the transmitters themselves. This uncertainty can lead to packet collisions, which occur whenever the underlying physical channel cannot support the transmission rates of all active users simultaneously. However, our viewpoint is that the  random level of the interference created by the random set of transmitters can also be exploited opportunistically by allowing transmission of different data streams, each of which might be decoded or not, depending on the  interference level at the receiver.

To be fair, the idea of transmitting information in layers in random access communication is not new; however an information-theoretic perspective of this layering idea was never  exposed.  Previously, Medard \textit{et al.}~\cite{Medard&Goldsmith:wireless04} studied the performance of Gaussian superposition coding in a  two-user additive white Gaussian noise (AWGN) system, but did not investigate the information-theoretic optimality of such a scheme. In the present work, we present coding schemes with guaranteed gaps to the information-theoretic capacity. We do so  under different channel models, and also extending the treatment to networks with a large number of users. Interestingly, it turns out that in the symmetric case in which all users are subject to the same received power constraint and are active with the same probability, superposition coding is not needed to achieve up to one bit from the throughput of an AWGN system.

The paper is organized in incremental steps, the first ones laying the foundation for the more complex scenarios.
Initially, we consider a two-user random access system, in which each sender can be in two modes of operation, active or not active. The set of active users is available to the decoder only, and active users encode data into two streams: one high priority stream ensures that part of the transmitted information is always
received reliably, while one low priority stream opportunistically takes advantage of the channel when the other user
is not transmitting.  Given this set-up, we consider two different channel models. First, we consider a binary-expansion deterministic (BD) channel model in which the input symbols are bits  and the output is the binary sum of a shifted version of the codewords sent by the transmitters. This is a first-order approximation of an AWGN channel in which the  shift of each input sequence corresponds to the amount of path loss experienced by the communication link. In this case, we exactly characterize the capacity region and it turns out that senders need to simultaneously transmit both streams to achieve capacity. Second,  we consider the AWGN channel and present a coding scheme
 that combines time-sharing and Gaussian superposition coding. This turns out to be within $\sqrt{3}/2$ bit from capacity. Furthermore, we also show that in the symmetric case in which both users are subject to the same received power constraint, superposition coding is not needed to achieve up to $\sqrt{3}/2$ bit from capacity.

Next, we consider  an $m$-user random access system, in which active transmitters encode data into independent streams of information, a subset of which are decoded by a common receiver, depending on the value of the collective interference. We cast this communication problem into an equivalent information-theoretic network with
multiple transmitters and receivers and we focus on the  \emph{symmetric} scenario in which users are active with the same probability $p$, independently of each other, and are subject to the same received power constraint, and we study the maximum achievable expected sum-rate ---videlicet throughput.
Given this set-up, we again consider the two channel models described above. First, we consider the BD channel model in the symmetric case in which all codewords are shifted by the same amount. In this setting, input and output symbols are bits, so that the receiver observes the binary sum of the codewords sent by the active transmitters. The possibility of decoding different messages in the event of multiple simultaneous transmissions depends on the rate at which the transmitted messages were encoded. Colliding codewords are correctly decoded when the sum of the rates at which they were encoded does not exceed one. This is a natural generalization of the classic packet collision model widely used in the networking literature, where packets are always encoded at rate one, so that transmissions are successful only when there is one active user.
We present a simple coding scheme which does \emph{not} employ superposition coding and which achieves the throughput. The coding scheme can be described as follows. When $p$ is close to zero, active transmitters ignore the presence of potential interferers and transmit a stream of data encoded at rate equal to one. By doing so, decoding at the receiver is successful if there is only one active user, and it fails otherwise.  This is what happens in  the classic slotted ALOHA protocol, for which a collision occurs whenever two or more users are simultaneously active in a given slot. In contrast, when $p$ is close to one, the communication channel is well approximated by the standard $m$-user binary sum DM-MAC, for which the number of transmitters is fixed and equal to $m$. In this regime, active users transmit a stream of data encoded at rate equal to $\tfrac{1}{m}$, that is, each active user requests an equal fraction of the $m$-user binary sum  DM-MAC sum-rate capacity. Any further increase in the per-user encoding rate would result in a collision. When $p$ is not close to either of the two extreme values, based on the total number of users $m$ and the access probability $p$, transmitters estimate the number of active users by solving a set of polynomial equations. If $k$ is the inferred number, then transmitters send one stream of data encoded at rate $\tfrac{1}{k}$, that is, each user requests an equal fraction of the $k$-user binary sum DM-MAC sum-rate capacity. Interestingly, it turns out that the estimator needed to achieve the throughput is different from the maximum-likelihood estimator $\lfloor mp \rfloor$ for the number of active users.
 The analysis also shows that the performance of slotted ALOHA systems can be improved by allowing \textit{encoding rate adaptation} at the transmitters. In fact, we show that the expected sum-rate of our proposed scheme tends to one as $m$ tends to infinity. Hence, there is no loss due to packet collisions  in the so called scaling limit of large networks. This  is in striking contrast with the well known behavior of slotted ALOHA systems in which users cannot adjust the encoding rate, for which the expected sum-rate tends to zero as $m$ tends to infinity. In practice, however, medium access schemes such as 802.11x typically use backoff mechanisms to effectively adapt the rates of the different users to the channel state. It is interesting to note that while these rate control strategies used in practice are similar to the information-theoretic optimum scheme described above for the case of equal received powers, practical receivers typically implement suboptimal decoding strategies, such as decoding one user while treating interference as noise.

Next, we consider  the case of the $m$-user AWGN channel. For this channel, we present a simple coding scheme which does not employ superposition coding  and which achieves the throughput to within one bit --- for any value of the underlying parameters. Perhaps not surprisingly, this coding scheme is very similar to the one described above for the case of the BD channel. In fact, the close connection between these two channel models has recently been exploited  to solve capacity problems for AWGN networks through their deterministic model counterpart \cite{Salman}.

Finally, we wish to mention some additional related works.
Extensions of ALOHA   resorting to probabilistic models to explain when multiple packets can be decoded in the presence of other simultaneous transmissions appear in \cite{GhezVerdu:AC88} and \cite{NawareMergerTong:IT05}.
An information-theoretic model to study layered coding in a two-user AWGN-MAC with no channel state information (CSI) available to
the transmitters was presented in a preliminary incarnation of this work~\cite{Minero:ISIT07}.  The two-user BD channel has been studied in  the adaptive capacity framework  in~\cite{elgamal}  and in this paper we also provide a direct comparison with that model.  We also rely on the  broadcast approach which has been pursued in \cite{Shamai:ISIT00}, and \cite{Shamai&Steiner:IT03} to study multiple access channels with no CSI available. A survey of the broadcast approach and its application to the analysis of multiple antenna systems appeared in~\cite{Shamai&Steiner:2008}, and we refer the reader to this work and to \cite{Biglieri} for an overview of the method and for additional references. The DM-MAC with partial CSI was studied in \cite{Steinberg} assuming two compressed descriptions of the state are available to the encoders.

The rest of the paper is organized as follows.
The next section formally defines the problem in the case of a two-user AWGN random access system. Section~\ref{sec:m} presents the extension to of the $m$-user random access system assuming an additive channel model. Section~\ref{sec:mB} consider the case of a BD channel model, while section~\ref{sec:mG} deals with the AWGN channel. A discussion about practical considerations and limitations of our model concludes the paper.

\section{The two-user Additive Random Access Channel}
\label{sec:m}
\begin{figure}
\begin{center}
\scalebox{1}{
\psfrag{x1}{ {$\bb{X}_1$}}
\psfrag{x2}{ {$\bb{X}_2$}}
\psfrag{z}{ {$\bb{Z}$}}
\psfrag{y}{ {$\bb{Y}$}}
\psfrag{W1}{ }
\psfrag{W2}{ }
\psfrag{Tx1}{  $\text{Tx}_1$}
\psfrag{Tx2}{  $\text{Tx}_2$}
\psfrag{Rx}{  $\text{Rx}$}
\psfrag{s1}{  $S_1$}
\psfrag{s2}{  $S_2$}
\psfrag{t1}{  $S_1 \in \{0,1\}$}
\psfrag{t2}{  $S_2 \in \{0,1\}$}
\includegraphics[width=3.7in]{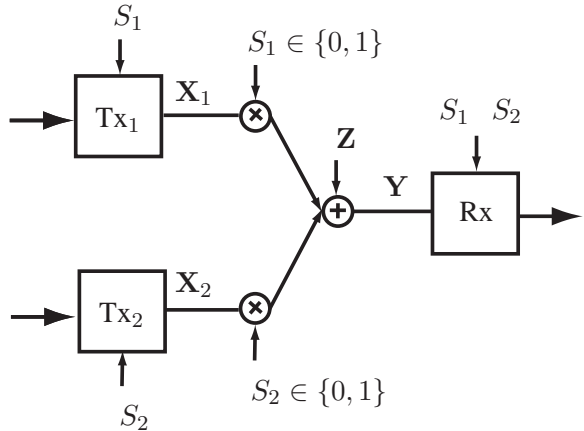}}
\end{center}
\caption{The two-user MAC with partial CSI modeling random access communications. The variables in bold represent vectors of length $n$.} \label{fig:1}
\end{figure}

Consider a two-user synchronous additive DM-MAC where each sender can be in two modes of operation, active or not active, independently of each other. The set of active users is available to the decoder, while encoders only know their own mode of operation. This problem is the compound DM-MAC with distributed state information depicted in Fig.~\ref{fig:1}. Specifically, the state of the channel is determined by two statistically independent binary random variables $S_1$ and $S_2$, which indicate whether user one and user two, respectively, are active, and it remains unchanged during the course of a transmission. Each sender knows its own state, while the receiver knows all the senders' states.
The presence of side information allows each transmitter to adapt its coding scheme to its state component. We can assume without loss of generality that senders transmit a codeword only when active, otherwise they remain silent.

Each sender transmits several streams of data, which are modeled via independent information messages, a subset of which is decoded by the common receiver, depending on the state of the channel. The notation we use is as follows. We denote by $\mathcal{W}_1=\{W_{1,1},\dotsc,W_{1,|\mathcal{W}_1|}\}$ and $\mathcal{W}_2=\{W_{2,1},\dotsc,W_{2,|\mathcal{W}_2|}\}$  the ensemble of independent messages transmitted by user $1$ and user $2$, respectively. We assume that each message $W_{i,j}$ is a random variable independent of everything else and uniformly distributed over a set with cardinality $2^{n R_{i,j}}$, for some non-negative rate $R_{i,j}$, $j \in \{1, \dotsc, |\mathcal{W}_i| \}$, $i\in\{1,2\}$.
We let $\mathcal{W}_i(A) \subseteq \mathcal{W}_i$ denotes the set of messages transmitted by user $i$, $i\in\{1,2\}$, that are decoded when the set of senders $A \subseteq \{1,2\}$ is active. Finally $r_i(A)$ denotes the sum of the rates at which messages in $\mathcal{W}_i(A)$ are encoded.

\begin{figure}
\begin{center}
\scalebox{.95}{
\psfrag{x1}{ {$\bb{X}_1$}}
\psfrag{x2}{ {$\bb{X}_2$}}
\psfrag{z1}{ {$\bb{Z}_1$}}
\psfrag{z2}{ {$\bb{Z}_2$}}
\psfrag{z12}{ {$\bb{Z}_{12}$}}
\psfrag{y1}{ {$\bb{Y}_1$}}
\psfrag{y2}{ {$\bb{Y}_2$}}
\psfrag{y12}{ {$\bb{Y}_{12}$}}
\psfrag{z0}{ {$Z_{\emptyset}$}}
\psfrag{y0}{ {$Y_{\emptyset}$}}
\psfrag{W1}{ $\mathcal{W}_1$}
\psfrag{W2}{ $\mathcal{W}_2$}
\psfrag{Wp1}{ $\mathcal{W}_1(\{1\})$}
\psfrag{Wp2}{ $\mathcal{W}_2(\{2\})$}
\psfrag{Wm1}{ $\mathcal{W}_1(\{1,2\})$}
\psfrag{Wm2}{ $\mathcal{W}_2(\{1,2\})$}
\psfrag{Tx1}{  $\text{Tx}_1$}
\psfrag{Tx2}{  $\text{Tx}_2$}
\psfrag{Rx1}{  $\text{Rx}_1$}
\psfrag{Rx2}{  $\text{Rx}_2$}
\psfrag{Rx12}{  $\text{Rx}_{12}$}
\psfrag{N}{  $\mathcal{N}(0,1)$}
\psfrag{C}{  $\C(\Power)$}
\includegraphics[width=3.7in]{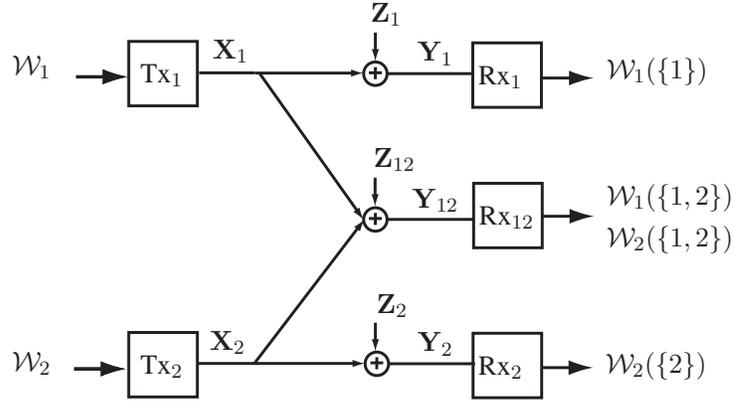}}
\end{center}
\caption{The network used to model a two-user random access system. The subscript index in $\bb{Y}$ and in $\bb{Z}$ denote the set of active users,  $\mathcal{W}_1(\{1,2\})$ and $\mathcal{W}_1(\{1,2\})$ encode information which is always decoded, while $\mathcal{W}_1(\{1\}) \setminus \mathcal{W}_1(\{1,2\})$ and $\mathcal{W}_2(\{2\}) \setminus \mathcal{W}_2(\{1,2\})$ denote messages which are decoded when there is no interference.} \label{fig:1b}
\end{figure}

Therefore, we can distinguish three non-trivial cases: if user 1 is the only active user, then the receiver decodes the messages in $\mathcal{W}_1(\{1\})$ and the transmission rate is equal to $r_1(\{1\})$; similarly, if user 2 is the only active user, then the receiver decodes the messages in $\mathcal{W}_2(\{2\})$, which are encoded at total rate of $r_2(\{2\})$;
finally, the receiver decodes messages in $\mathcal{W}_1(\{1,2\})$ and $\mathcal{W}_2(\{1,2\})$ when both users are active, so senders communicate at rate $r_1(\{1,2\})$ and $r_2(\{1,2\})$, respectively. The resulting information-theoretic network is illustrated in Fig.~\ref{fig:1b}, where one auxiliary receiver is introduced for each channel state component. It is clear from the figure that, upon transmission, each transmitter is connected to the receiver either through a point-to-point link or through an additive DM-MAC, depending on the channel state.

Observe that if the additive noises in Fig.~\ref{fig:1b} have the same
marginal distribution, then the channel output sequence observed by
the MAC receiver is a \emph{degraded} version of the sequence observed
by each of the two point-to-point receivers, because of the mutual
interference between the transmitted codewords. As in a degraded
broadcast channel, the ``better" receiver can always decode the
message intended for the ``worse'' receiver, similarly here each
point-to-point receiver can decode what can be decoded at the MAC receiver. Thus, there is
no loss of generality in assuming that \begin{equation}
\label{eq:assumption1}
\mathcal{W}_1(\{1,2\}) \subseteq \mathcal{W}_1(\{1\})
\end{equation}
and that
\begin{equation}
\label{eq:assumption2}
\mathcal{W}_2(\{1,2\}) \subseteq \mathcal{W}_2(\{2\}).
\end{equation}
Then, messages in $\mathcal{W}_1(\{1,2\})$ and $\mathcal{W}_1(\{1,2\})$  ensure that some transmitted information is
always received reliably, while the remaining messages provide additional information that can be opportunistically
decoded when there is no interference.
If conditions \eqref{eq:assumption1} and \eqref{eq:assumption2} are satisfied, then we say that $\mathscr{W}=(\{\mathcal{W}_1,\mathcal{W}_2\},\{ \mathcal{W}_1(\{1\}),\mathcal{W}_1(\{1,2\}),\mathcal{W}_2(\{2\}),\mathcal{W}_2(\{1,2\}) \})$ denotes a \emph{message structure} for the channel in Fig.~\ref{fig:1b}.

For a given message structure $\mathscr{W}$, we say that the rate tuple $(r_1(\{1\}),r_2(\{2\}),r_1(\{1,2\}),r_2(\{1,2\}))$ is achievable if there exist a sequence of coding and decoding functions such that each receiver in Fig.~\ref{fig:1b} can decode all intended messages with arbitrarily small error probability as the coding block size tends to infinity. We define the capacity region $\Ca{\mathscr{W}}$ as the closure of the set of achievable rate tuples.

Observe that as we vary $|\mathcal{W}_1|$, $|\mathcal{W}_2|$, and the sets of decoded messages, there are infinitely many possible message structures for a given channel. For each one of them we define $\Ca{\mathscr{W}}$.

Next, we define the \emph{capacity} of the channel in Fig.~\ref{fig:1b}, denoted by $\Ca{}$, as the closure of the union of $\Ca{\mathscr{W}}$ over all possible message structures $\mathscr{W}$. Note that $\Ca{}$ represents the optimal tradeoff among the rates $(r_1(\{1\}),$ $r_2(\{2\}),$ $r_1(\{1,2\}),$ $r_2(\{1,2\})$ over \emph{all} possible ways of partitioning information into different information messages such that conditions \eqref{eq:assumption1} and \eqref{eq:assumption2} are satisfied.


In the next section we answer the question of characterizing $\Ca{}$ for two additive channels of practical interest. First, we consider the BD channel model, for which we completely characterize the capacity region $\Ca{}$. Perhaps not surprisingly, we show that to achieve $\Ca{}$ it suffices that each sender  transmits \emph{two} independent information messages, one of which carries some reliable information which is always decoded, while the remaining one carries additional information which is decoded when the other user is not transmitting. Second, we consider the AWGN channel, for which we provide a constant gap characterization of $\Ca{}$, where the constant is universal and independent of the channel parameters.
Finally, we apply this result to the study of the throughput of a two-user random access system under symmetry assumptions.

\section{Example 1: the two-user BD random access channel}
\label{sec:2d}
\begin{figure}
\begin{center}
\scalebox{1}{
\psfrag{x1}{ {$X_1$}}
\psfrag{x2}{ {$X_2$}}
\psfrag{z}{ {$Z$}}
\psfrag{y}{ {$Y$}}
\psfrag{W11}{ $\wwp{1}$}
\psfrag{W21}{ $\wwp{2}$}
\psfrag{W12}{ $\wwc{1}$}
\psfrag{W22}{ $\wwc{2}$}
\psfrag{WW11}{ $\hwp{1}$}
\psfrag{WW21}{ $\hwp{2}$}
\psfrag{WW12}{ $\hwc{1}$}
\psfrag{WW22}{ $\hwc{2}$}
\psfrag{Tx1}{  $\text{Tx}_1$}
\psfrag{Tx2}{  $\text{Tx}_2$}
\psfrag{Rx1}{  $\text{Rx}_1$}
\psfrag{Rx2}{  $\text{Rx}_2$}
\psfrag{Rx12}{  $\text{Rx}_{12}$}
\psfrag{r1}{  $n_1$}
\psfrag{r2}{  $n_2$}
\psfrag{t1}{  $S_1 \in \{0,1\}$}
\psfrag{t2}{  $S_2 \in \{0,1\}$}
\includegraphics[width=3.7in]{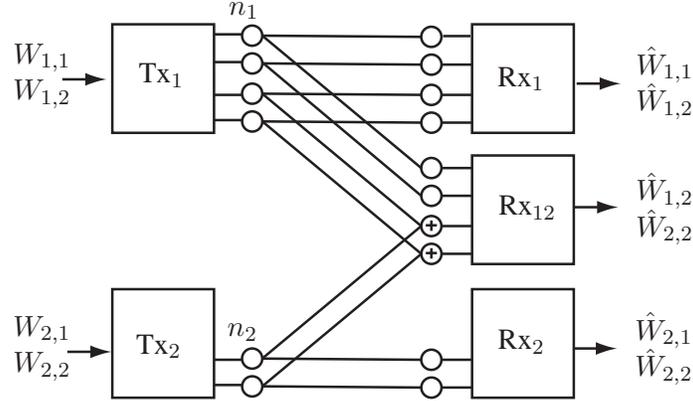}}
\end{center}
\caption{The two-user BD-RAC, and the message structure used to prove the achievability of the capacity region.} \label{fig:1d}
\end{figure}

Suppose that channel input and output alphabets are each the set $\{0,1\}^{n_1}$, for some integer number $n_1$, and that at each time unit $t\in\{1,\dotsc,n\}$ inputs and outputs are related as follows:
\begin{equation}
\label{eq:2det}
\setlength\arraycolsep{0.2em}
  \begin{array}{lcl}
Y_{1,t} & =&  X_{1,t}, \\
Y_{12 ,t} & =& X_{1,t} + S^{n_1-n_2} X_{2,t}, \\
Y_{2 ,t} & =& S^{n_1-n_2} X_{2,t},
\end{array}
\end{equation}
where $n_2 \le n_1$ denotes an integer number, summation and product are over $\text{GF}(2)$, and $S^{n_1-n_2}$ denotes the $(n_1-n_2) \times (n_1-n_2)$ shift matrix having the $(i,j)$th component equal to $1$ if $i = j + (n_1-n_2)$, and $0$ otherwise. By pre-multiplying $X_{2,t}$ by $S^{n_1-n_2}$, the first $n_2$ components of $X_{2,t}$ are down-shifted by $(n_1-n_2)$ positions and the remaining elements are  set equal to zero. We refer to this model as the two-user BD \textit{random access channel} (RAC), see Fig.~\ref{fig:1d} for a pictorial representation. Physically, this channel represents a first-order approximation of a wireless channel in which continuous signals are represented by their binary expansion, the codeword length $n_1$ represents the noise cut-off value, and the amount of shift $n_1-n_2$ corresponds to the path loss of user 2 relative to use 1 \cite{Salman}.
The following theorem characterizes the capacity region of this channel.
\begin{thm}
\label{thm:1}
The capacity region $\Ca{}$ of the two-user BD-RAC is the set of non-negative rate tuples such that
\begin{equation}
\setlength\arraycolsep{0.2em}
  \begin{array}{rcl}
r_1(\{1\}) &\leq & n_1,  \\
r_2(\{2\})  &\leq & n_2, \\
r_1(\{1\}) + r_2(\{1,2\}) & \leq & n_1, \\
r_2(\{2\}) + r_1(\{1,2\}) & \leq & n_1, \\
r_1(\{1,2\})  & \leq & r_1(\{1\}), \\
r_2(\{1,2\})  & \leq & r_2(\{2\}).
\end{array}
\label{eq:in_det}
\end{equation}
\end{thm}

The proof of the converse part of the above theorem can be sketched as follows. Observe that the common receiver observing $\bb{Y}_{12} \triangleq \{ Y_{12,1},\dotsc,Y_{12,n} \}$ can decode messages in $\mathcal{W}_2(\{1,2\})$. Let us suppose that this receiver is given messages in $\mathcal{W}_2 \setminus \mathcal{W}_2(\{1,2\})$ as side information. Then, it has full knowledge of $\mathcal{W}_2 $, so it can compute the codeword $\bb{X}_2$ transmitted by user 2, subtract it from the aggregate received signal $\bb{Y}_{12}$, obtaining $\bb{X}_1$. Thus, given the side information, the channel output observed by the common receiver becomes statistically equivalent to $\bb{Y}_1$. Since receiver $1$ can decode $\mathcal{W}_1(\{1\})$ upon observing $\bb{Y}_1$, we conclude that receiver $12$ must also be able to decode message $\mathcal{W}_2(\{1,2\})$. Hence, $ r_1(\{1\}) + r_2(\{1,2\}) \le n_1$.
By providing side information about message $\mathcal{W}_1 \setminus \mathcal{W}_1(\{1,2\})$ and following the same argument above, we obtain that $ r_2(\{2\}) + r_1(\{1,2\}) \le n_1$. The remaining bounds are trivial.

The proof of the achievability part of the theorem shows that it suffices to partition information into \emph{two} independent messages, such that $\mathcal{W}_1=\{\wwp{1},\wwc{1} \}$ and $\mathcal{W}_2=\{\wwp{2},\wwc{2} \}$. Messages $\wwc{1}$ and $\wwc{2}$ represent ensure that part of the transmitted information is always
received reliably, while $\wwp{1}$  and $\wwp{2}$ are decoded opportunistically when one user is not transmitting. The corresponding message structure is illustrated in Fig.~\ref{fig:1d}.  In general, the coding scheme which we employ in the proof of the achievability requires that user 1 simultaneously transmits $\wwp{1}$ and $\wwc{1}$. However, in the special symmetric case in which $n_1=n_2$ all rate  tuples in the capacity region can be achieved by means of coding strategies in which each user transmits only one of the two messages.

\begin{IEEEproof}
First, we prove the converse part of the theorem. The first two inequalities which define $\Ca{}$ are standard point-to-point bounds which can be derived via standard techniques. To obtain the third inequality, observe that
by Fano's inequality we have that $H(\mathcal{W}_1(\{1,2\})|\bb{Y}_{12}) \leq n \epsilon_{n}$, $H(\mathcal{W}_2(\{1,2\})|\bb{Y}_{12}) \leq n \epsilon_{n}$, as well as $H(\mathcal{W}_i(i)|\bb{Y}_{i}) \leq n \epsilon_{n}$, where $\epsilon_{n}$ tend to zero as the block length $n$ tends to infinity. From the independence of the source messages, we have that
\begin{align}
n(r_1(\{1\}) + r_2(\{1,2\}) ) & = H(\mathcal{W}_1(\{1\}),\mathcal{W}_2(\{1,2\})), \nonumber  \\
& = H(\mathcal{W}_1(\{1\}),\mathcal{W}_2(\{1,2\}) |\mathcal{W}_2 \setminus \mathcal{W}_2(\{1,2\}) ), \nonumber  \\
& = I(\mathcal{W}_1(\{1\}),\mathcal{W}_2(\{1,2\});\bb{Y}_{12}| \mathcal{W}_2 \setminus \mathcal{W}_2(\{1,2\}) ), \nonumber \\
& \quad \quad +  H( \mathcal{W}_1(\{1\}),\mathcal{W}_2(\{1,2\})|\bb{Y}_{12}, \mathcal{W}_2 \setminus \mathcal{W}_2(\{1,2\}) ).
\label{eq:2t1}
\end{align}
Using the memoryless property of the channel and the fact that conditioning reduces the entropy, the first term in the right hand side of \eqref{eq:2t1} can be upper bounded as
\begin{align}
I(\mathcal{W}_1(\{1\}),\mathcal{W}_2(\{1,2\});\bb{Y}_{12}| \mathcal{W}_2 \setminus \mathcal{W}_2(\{1,2\}) ) \leq n n_1. \label{eq:2t2}
\end{align}
On the other hand, from the chain rule, the fact that conditioning reduces the entropy, and Fano's inequality, we have that
\begin{align}
& H( \mathcal{W}_1(\{1\}),\mathcal{W}_2(\{1,2\})|\bb{Y}_{12}, \mathcal{W}_2 \setminus \mathcal{W}_2(\{1,2\}) ) \nonumber  \\
& \; =  H( \mathcal{W}_2(\{1,2\})|\bb{Y}_{12}, \mathcal{W}_2 \setminus \mathcal{W}_2(\{1,2\}) ) + H( \mathcal{W}_1(\{1\})|\bb{Y}_{12}, \mathcal{W}_2) \nonumber  \\
& \; \le   H( \mathcal{W}_2(\{1,2\})|\bb{Y}_{12}) + H( \mathcal{W}_1(\{1\})|\bb{Y}_{12}, \mathcal{W}_2, \bb{X}_{2}) \nonumber  \\
& \; = H( \mathcal{W}_2(\{1,2\})|\bb{Y}_{12}) + H( \mathcal{W}_1(\{1\})|\bb{Y}_{1}) \nonumber  \\
& \;\leq 2n\epsilon_{n}  \label{eq:2t3}
\end{align}
where the last equality is obtained observing from \eqref{eq:2det} that, if $\bb{X}_{2}$ is given, then $\bb{Y}_{12}$ is statistically equivalent to $\bb{Y}_{1}$. Substituting \eqref{eq:2t2} and \eqref{eq:2t3} into \eqref{eq:2t1}, we have that
\begin{align*}
& n(r_1(\{1\}) + r_2(\{1,2\}) )   \le n n_1 + 2n\epsilon_{n},
\end{align*}
and the desired inequality is obtained in the limit of $n$ going to infinity. The fourth inequality in \eqref{eq:in_det} is obtained by a similar argument. Finally, the last two inequalities in \eqref{eq:in_det} follow from \eqref{eq:assumption1} and \eqref{eq:assumption2}.

Next, to prove the direct part of the theorem, we establish that $\Ca{}$ is equal to the capacity of the two-user BD-RAC for the specific message structure defined by $\mathcal{W}_i = \{\wwp{i},\wwc{i}\}$, $\mathcal{W}_i(\{i\}) = \mathcal{W}_i$, and $\mathcal{W}_i(\{12\}) = \{\wwc{i}\}$, $i \in \{1,2\}$.
For this message structure we have that
\begin{equation}
\setlength\arraycolsep{0.2em}
  \begin{array}{rcl}
r_1(\{1\}) &=&  \Rc{1} + \Rp{1}, \\
r_2(\{2\}) &= & \Rc{2} + \Rp{2}, \\
r_1(\{1,2\}) &=&\Rc{1}, \\
r_2(\{1,2\}) &=&\Rc{2}.
\end{array}
\label{eq:in_det2}
\end{equation}
We have established above that if $(r_1(\{1\}),$ $r_2(\{2\}),$ $r_1(\{1,2\}),$ $r_2(\{1,2\}) \in \Ca{\mathscr{W}}\subseteq \Ca{}$, then inequalities \eqref{eq:in_det} have to be satisfied. Combining the non-negativity of the rates, \eqref{eq:in_det}, and \eqref{eq:in_det2}, and eliminating $(r_1(\{1\}),$ $r_2(\{2\}),$ $r_1(\{1,2\}),$ $r_2(\{1,2\})$ from the resulting system of inequalities, we obtain
\begin{equation}
\setlength\arraycolsep{0.2em}
  \begin{array}{rcl}
\Rp{1} + \Rc{1} &\leq & n_1,  \\
\Rp{2} + \Rc{2}  &\leq & n_2, \\
\Rp{1} + \Rc{1} + \Rc{2} &\leq & n_1, \\
\Rp{2} + \Rc{1} + \Rc{2} &\leq & n_1. \\
\end{array}
\label{eq:in_det_a}
\end{equation}
The above system of inequalities is the image of \eqref{eq:in_det} under the linear map \eqref{eq:in_det2}. Since the map is invertible, proving the achievability of all rate tuples $(r_1(\{1\}),$ $r_2(\{2\}),$ $r_1(\{1,2\}),$ $r_2(\{1,2\})$ satisfying \eqref{eq:in_det} is equivalent to proving the achievability of all rate tuples $(\Rp{1},$ $\Rp{2},$ $\Rc{1},$ $\Rc{2})$ satisfying \eqref{eq:in_det_a}. It is tedious but simple to verify that the set of non-negative rate tuples satisfying \eqref{eq:in_det_a} is equal to the convex hull of ten extreme points, four of which are dominated by one of the remaining six. Given two vectors $\bb{u}$ and $\bb{v}$, we say that $\bb{u}$ \emph{dominates} $\bb{v}$ if each coordinate of $\bb{u}$
is greater than or equal to the corresponding coordinate of $\bb{v}$.
The six dominant extreme points of \eqref{eq:in_det_a} are given by
$\bb{v}_1=[n_2,n_2,n_1-n_2,0]^T$, $\bb{v}_2=[n_1-n_2,0,0,n_2]^T$, $\bb{v}_3=[0,0,n_1-n_2,n_1]^T$, $\bb{v}_4=[n_1,n_2,0,0]^T$, $\bb{v}_5=[0,0,n_1,0]^T$, $\bb{v}_6=[0,0,0,n_2]^T$, where the four coordinates denote $(\Rp{1},$ $\Rp{2},$ $\Rc{1},$ $\Rc{2})$, respectively.

The achievability of $\bb{v}_1,\dotsc,\bb{v}_6$ can be sketched as follows. To achieve $\bb{v}_1$ sender 1 transmit simultaneously $\wwc{1}$ and $\wwp{1}$, in the first $n_1-n_2$ and last  $n_2$ components of $X_1$, respectively. User 2, instead, transmits $\wwp{2}$ in the first $n_2$ components of $X_2$.  Because of the downshift in $X_2$, the multiple access decoder receives the binary sum of $\wwp{1}$ and $\wwp{2}$ in the last $n_2$ components of $Y_{12}$, and can successfully decoded $\wwc{1}$ from the first $n_1-n_2$ interference-free components. Effectively, coding is performed so that $\wwp{1}$ and $\wwp{2}$ are received ``aligned'' at the common receiver, see Fig.~\ref{fig:proof} for a pictorial representation. Observe that in the special case in which $n_1=n_2$, sender 1 only transmit message $\wwc{1}$. Likewise, $\bb{v}_2,\dotsc,\bb{v}_6$ can be achieved by transmitting one message per user, in such a way that the transmitted codewords do not interfere with each other at the multiple access receiver. For example, to achieve $\bb{v}_1$ user 2 transmits $\wwc{2}$ in the first $n_2$ components of $X_2$, while user 1 transmits $\wwc{1}$ in the first $n_1-n_2$ components of $X_1$.

Next, observe that if an extreme point $\bb{v}$ is achievable, then all extreme points dominated by $\bb{v}$ are also achievable by simply decreasing the rate of some of the messages. Finally, any point in \eqref{eq:in_det} can be written as convex combination of the extreme points, hence it can be achieved by time-sharing among the basic coding strategies which achieve $\bb{v}_1,\dotsc,\bb{v}_6$. This shows that all rate tuples satisfying \eqref{eq:in_det_a} are achievable.
\begin{figure}
\begin{center}
\scalebox{1}{
\psfrag{x1}{ {$X_1$}}
\psfrag{x2}{ {$X_2$}}
\psfrag{z}{ {$Z$}}
\psfrag{y}{ {$Y$}}
\psfrag{W11}{ $\wwp{1}$}
\psfrag{W21}{ $\wwp{2}$}
\psfrag{W12}{ $\wwc{1}$}
\psfrag{W22}{ $\wwc{2}$}
\psfrag{WW11}{ $\hwp{1}$}
\psfrag{WW21}{ $\hwp{2}$}
\psfrag{WW12}{ $\hwc{1}$}
\psfrag{WW22}{ $\hwc{2}$}
\psfrag{Tx1}{  $\text{Tx}_1$}
\psfrag{Tx2}{  $\text{Tx}_2$}
\psfrag{Rx1}{  $\text{Rx}_1$}
\psfrag{Rx2}{  $\text{Rx}_2$}
\psfrag{Rx12}{  $\text{Rx}_{12}$}
\psfrag{r1}{  $n_1$}
\psfrag{r2}{  $n_2$}
\psfrag{r3}{  $n_1-n_2$}
\psfrag{t1}{  $S_1 \in \{0,1\}$}
\psfrag{t2}{  $S_2 \in \{0,1\}$}
\includegraphics[width=3.7in]{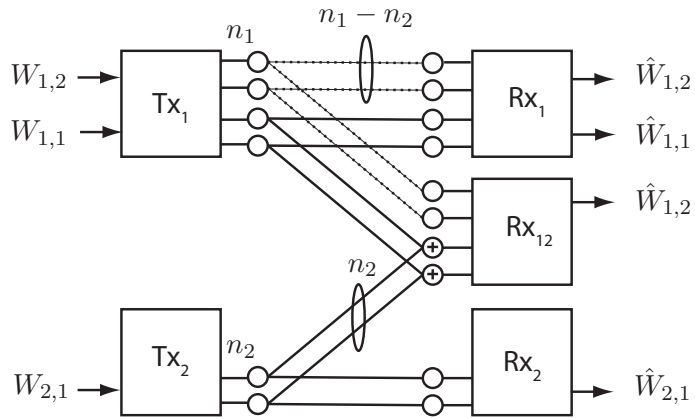}}
\end{center}
\caption{Pictorial representation of the coding scheme which achieves the rate tuple $R_{1,1}=n_2$, $R_{1,2}=n_1-n_2$, $R_{2,1}=n_2$, $R_{1,2}=0$. Message $\wwp{2}$ is transmitted via the first $n_1-n_2$ interference-free links (dotted lines), while $\wwp{1}$ and $\wwp{2}$ are sent through the remaining $n_2$ links (solid lines), so that the interference they generate results ``aligned'' at the common receiver.} \label{fig:proof}
\end{figure}
\end{IEEEproof}

\subsection{The throughput in a symmetric scenario.}
\label{sec:t2}
Having an exact characterization of the capacity region at hands, it is now possible to formulate and solve optimization problems of practical interests. As an example, we consider the problem of maximizing the throughput in the symmetric scenario where $n_1 = n_2 = 1$, and where each user is independently active with probability $p$.

This model represents a first-order approximation of a wireless channel in which data arrivals follow the same law, and where transmitted signals are received at the same power level. The codeword length is normalized to 1 so that the maximum amount of information which can be conveyed across the channel is one bit per channel use, regardless the number of active users.
The possibility of decoding different messages in the event of multiple simultaneous transmissions depends on the rate at which the messages were encoded. Colliding codewords are correctly decoded when the sum of the rates at which they were encoded does not exceed one. This is a natural generalization of the classic packet collision model widely used in the networking literature, where packets are always encoded at rate one, so that transmissions are successful only when there is one active user.
The parameter $p$ represents the burstiness of data arrivals, and determines the law of the variables $S_1$ and $S_2$ in Fig.~\ref{fig:1}, hence the channel law. Based on the knowledge of $p$, each sender can ``guess'' the state of operation of the other user, and optimize the choice of the encoding rates so that the expected sum-rate, or throughput, is maximized.

Formally, we look for the solution of the following optimization problem:
\begin{equation*}
\max p(1-p)\left[ r_1(\{1\}) + r_2(\{2\}) \right] + p^2 \left[ r_1(\{1,2\}) + r_2(\{1,2\}) \right]
\end{equation*}
subject to the constraint that the rates should be in $\Ca{}$. Observe that the weight assigned to each rate component $r_i(A)$ is uniquely determined by $p$, and is equal to the probability that users in the set $A$ are active. By means of Theorem \ref{thm:1}, it is easy to show that the solution to the above problem is equal to
\begin{equation*}
\left\{
  \begin{array}{ll}
    2 p (1-p), & \hbox{if } p \in (0, 1/2]; \\
    p, & \hbox{if } p \in (1/2,1].
  \end{array}
\right.
\end{equation*}
The coding strategy used to achieve the throughput can be described as follows. If the transmission probability $p$ lies in the interval $(0, 1/2]$, then user $i$ transmits message $\wwp{i}$ encoded at rate 1. A collision occurs in the event that both senders are simultaneously transmitting, which occurs with probability $p^2$, in which case the common receiver cannot decode the transmitted codewords. Decoding is successful if only one of the two users is active, so the expected sum-rate achieved by this scheme is equal to $2 p (1-p)$. If, instead, the transmission probability $p$ lies in the interval $(1/2,1]$, then user $i$ transmits message $\wwc{i}$ encoded at rate $1/2$, i.e., at half the sum-rate capacity of the two-user binary additive MAC. By doing so, the transmitted codewords are never affected by collisions, and can be decoded in any channel state. This yields an expected sum-rate of $2 p (1-p) 1/2 + p^2 $.
It should be highlighted that in this symmetric scenario each user transmits only one of the two messages for any value of $p$.

We show later in the paper that this optimization problem can be solved in the general case of a network with more than two users.

\section{Example 2: the two-user AWGN-RAC}
\label{sec:2}
We now turn to another example of additive channels.  Assume that at each discrete time step inputs and outputs are related as follows:
\begin{equation}
\label{eq:2gau}
\setlength\arraycolsep{0.2em}
  \begin{array}{lcl}
Y_{1,t} & = & X_{1,t} + Z_{1,t}, \\
Y_{12 ,t} & = & X_{1,t} + X_{2,t} + Z_{12,t}, \\
Y_{2 ,t} & = & X_{2,t} + Z_{2,t},
\end{array}
\end{equation}
where $Z_{1,t}$, $Z_{2,t}$, and $Z_{12,t}$ are independent standard Gaussian random variables, and the sum is over the field of real numbers.
Assume that the realizations of $\{X_{i,t}\}$ satisfy the following average power constraint
\begin{equation*}
\sum_{t=1}^n x_{i,t}^2 \leq n\Power_i
\end{equation*}
for some positive constant $\Power_i$, $i=1,2$, and that $\Power_1 \geq \Power_2$. We refer to the model in \eqref{eq:2gau} as the two-user AWGN-RAC. In the rest of the paper, we use the notation $\C(x) \triangleq 1/2 \log(1+x)$.

An outer bound to the capacity region $\Ca{}$ of the two-user AWGN-RAC in \eqref{eq:2gau} is given by the following Theorem.
\begin{thm}
\label{thm:out2f}
Let $\Ro{}$ denote the set of non-negative rates such that
\begin{equation}
\setlength\arraycolsep{0.2em}
  \begin{array}{rcl}
r_1(\{1\}) & \leq &  \C(\Power_1),  \\
r_2(\{2\}) & \leq &  \C(\Power_2), \\
r_1(\{1\}) + r_2(\{1,2\})  & \leq &  \C(\Power_1 + \Power_2), \\
r_2(\{2\}) + r_1(\{1,2\}) & \leq &  \C(\Power_1 + \Power_2), \\
r_1(\{1,2\})  & \leq & r_1(\{1\}), \\
r_2(\{1,2\})  & \leq & r_2(\{2\}).
\end{array}
\label{eq:in_gau}
\end{equation}
Then, $\Ca{} \subseteq \Ro{}$.
\mbox{}
\end{thm}

The proof of the above theorem is similar to the converse part of Theorem \ref{thm:1} and it is hence omitted.

Next, we prove an achievability result by computing an inner bound to the capacity region $\Ca{\mathscr{W}}$ of the two-user AWGN-RAC for a specific message structure $\mathscr{W}$. As for the BD-RAC, we let $\mathcal{W}_i = \{\wwp{i},\wwc{i}\}$, $\mathcal{W}_i(i) = \mathcal{W}_i$, and $\mathcal{W}_i(12) = \{\wwc{i}\}$, $i \in \{1,2\}$.
The encoding scheme we use is Gaussian superposition coding. Each sender encodes the messages using independent Gaussian codewords having sum-power less or equal to the power constraint. Decoding is performed using successive interference cancelation: messages are decoded in a prescribed decoding order, treating interference of messages which follow in the order as noise. Then, each decoded codeword is subtracted from the aggregate received signal.

%
\begin{prop}
\label{thm:in2f}
Let $\Ri{\mathscr{W}}'$ denote the set of non-negative rates such that
\begin{equation}
\setlength\arraycolsep{0.2em}
  \begin{array}{rcl}
r_1(\{1\}) & \leq &  \C(\Power_1),  \\
r_2(\{2\})  & \leq &  \C(\Power_2), \\
r_1(\{1\}) + r_2(\{2\}) & \leq &  \C(\Power_1+\Power_2), \\
r_1(\{1,2\})  & \leq & r_1(\{1\}), \\
r_2(\{1,2\})  & = & r_2(\{2\}).
\end{array}
\label{eq:in_gau1}
\end{equation}
Similarly, let $\Ri{\mathscr{W}}''$ denote the set of non-negative rates satisfying \eqref{eq:in_gau1} after after swapping the indices 1 and 2.
Finally, let $\Ri{\mathscr{W}}'''$ denote the set of non-negative rates satisfying the following inequalities
\begin{equation}
\label{eq:in3}
\setlength\arraycolsep{0.2em}
  \begin{array}{rcl}
r_1(\{1,2\}) & \leq &  \C\left( \frac{(1-\beta_1)\Power_1}{\beta_1 \Power_1 + \beta_2 \Power_2 + 1} \right), \\
r_2(\{1,2\})  & \leq & \C\left( \frac{(1-\beta_2)\Power_2}{\beta_2 \Power_2 + \beta_2 \Power_2 + 1} \right), \\
r_1(\{1,2\})+r_2(\{1,2\})  & \leq & \C\left( \frac{(1-\beta_1)\Power_1+(1-\beta_2)\Power_2}{\beta_2 \Power_2 + \beta_2 \Power_2 + 1} \right), \\
r_1(\{1\})  & \leq &  r_1(\{1,2\}) + \C(\beta_1 \Power_1), \\
r_2(\{2\}) & \leq &  r_2(\{1,2\}) + \C(\beta_2 \Power_2).
  \end{array}
\end{equation}
for some $(\beta_1,\beta_2) \in [0,1]\times [0,1]$. Let $\Ri{\mathscr{W}} = \text{closure}(\Ri{\mathscr{W}}' \cup \Ri{\mathscr{W}}'' \cup \Ri{\mathscr{W}}''')$.
Then, $\Ri{\mathscr{W}} \subseteq \Ca{\mathscr{W}} \subseteq \Ca{} $.
\mbox{}
\end{prop}
\begin{proof}
Suppose that sender two does not transmit message $\wwp{2}$, i.e., $\Rp{2} = 0$. The achievability of $\Ri{2}'$ can then be shown by using a standard random coding argument as for the AWGN-MAC. To send $(\wwc{1},\wwp{1})$, encoder one sends the sum of two independent Gaussian codewords having sum-power equal to $\Power_1$. On the other hand, sender two encodes $\wwc{2}$ into a Gaussian codeword having power $\Power_2$. A key observation is that the common receiver observing $Y_{12}$ can decode all transmitted messages: $\wwc{1}$, $\wwc{2}$ can be decoded by assumption, while $\wwp{1}$ can be decoded after having subtracted $\bb{X}_2$ from the received channel output. Thus, by joint typical decoding, decoding is successful with arbitrarily small error probability if $ \Rp{1} +\Rc{1} +\Rc{2} < \C(\Power_1 + \Power_2)$, i.e., $r_1(\{1\}) + r_2(\{2\}) <  \C(\Power_1+\Power_2)$. Similarly, the receiver observing $Y_{1}$ can decode messages $\wwc{1}$, $\wwp{1}$ as long as $ \Rp{1} +\Rc{1} < \C(\Power_1)$, i.e., $r_1(\{1\}) <  \C(\Power_1)$ while the receiver observing $Y_{2}$ can decode messages $\wwc{2}$ if $r_2(\{2\}) \leq  \C(\Power_2)$. We conclude that $\Ri{2}'$ is an inner bound to the capacity region.
By swapping the role of user 1 and user 2 it is easy to see that $\Ri{2}''$ is also an inner bound to the capacity region.
We claim that $\Ri{2}'''$ can be achieved by a coding scheme which combines Gaussian superposition coding and multiple access decoding. As in the Gaussian broadcast channel, to send the message pair $\bigl(\wwp{i},\wwc{i} \bigr)$, encoder $i$ sends the codeword $\bb{X}_i \bigl( \wwp{i},\wwc{i} \bigr) = \bb{U}_i\bigl(\wwc{i} \bigr) + \bb{V}_i\bigl(\wwp{i} \bigr)$, where the sequences $\bb{U}_i$ and $\bb{V}_i$ are independent Gaussian codewords having power $(1-\beta_i) \Power_i$ and $\beta_i \Power_i $ respectively, $i=1,2$. Upon receiving $Y_{12}$, decoder 12 first decodes $\wwc{1} $ and $\wwc{2} $ using a MAC decoder and treating $\bb{V}_1\bigl(\wwp{1} \bigr)+\bb{V}_2\bigl(\wwp{2} \bigr)$ as noise. Decoding is successful with arbitrarily small error probability if
\begin{equation}
\label{eq:in3}
\setlength\arraycolsep{0.2em}
  \begin{array}{rcl}
\Rc{2} & < &  \C\left( \frac{(1-\beta_1)\Power_1}{\beta_1 \Power_1 + \beta_2 \Power_2 + 1} \right), \\
\Rc{1}  & < & \C\left( \frac{(1-\beta_2)\Power_2}{\beta_1 \Power_1 + \beta_2 \Power_2 + 1} \right), \\
\Rc{1}+\Rc{2}  & < & \C\left( \frac{(1-\beta_1)\Power_1+(1-\beta_2)\Power_2}{\beta_1 \Power_1 + \beta_2 \Power_2 + 1} \right).
  \end{array}
\end{equation}
Upon receiving $Y_{i}= \bb{U}_i\bigl(\wwc{i} \bigr) + \bb{V}\bigl(\wwp{i} \bigr) + \bb{Z}_{i}$, decoder $i$ performs decoding via successive interference cancelation: it first decodes $\wwc{i} $ treating $\bb{V}_i\bigl(\wwp{i} \bigr) + \bb{Z}_{i}$ as noise, then it subtracts $\bb{U}_i\bigl(\wwc{i} \bigr)$ from $Y_{i}$ and decodes $\wwp{i}$
 from $\bb{V}_i\bigl(\wwp{i} \bigr) + \bb{Z}_{i}$. Thus, decoding of $\wwc{i}$ is successful if $\Rc{i} < \C\bigl( \frac{(1-\beta_i) \Power_i}{\beta_i \Power_i +1 } \bigr)$, while decoding of $\wwp{i}$ is successful if $\Rp{i} < \C\bigl( \beta_1 \Power\bigr)$.
After combining these conditions to the equalities which relate $(r_1(\{1\}),$ $r_2(\{2\}),$ $r_1(\{1,2\}),$ $r_2(\{1,2\})$ to $(\Rp{1},$ $\Rp{2},$ $\Rc{1},$ $\Rc{2})$, and eliminating $(\Rp{1},$ $\Rp{2},$ $\Rc{1},$ $\Rc{2})$ from the resulting system of inequalities, we obtain that \eqref{eq:in3} have to be satisfied for the above coding scheme to work. Finally, a standard time-sharing argument can be used to show that the $\text{closure}(\Ri{2}' \cup \Ri{2}'' \cup \Ri{2}''') \subseteq \Ca{2} $
\end{proof}
The following theorem explicitly characterizes the gap between the above inner and outer bounds on $\Ca{}$.
\begin{thm}
\label{thm:2b}
Let $\bb{R} \in \Ro{}$. Then, there exists $\bb{R}' \in \Ri{\mathscr{W}}$ such that $\parallel \bb{R}  - \bb{R}' \parallel \le \frac{\sqrt{3}}{2}$.
\mbox{}
\end{thm}
\begin{proof}
See Appendix I.
\end{proof}

Observe Gaussian superposition coding is \emph{not} the optimal coding strategy for the AWGN channel under consideration. However, the above theorem ensures that Gaussian superposition coding achieves to within $\sqrt{3}/2$ bit from the capacity $\Ca{}$. It is important to note that this bound holds independently of the power constraints $\Power_1$ and $\Power_2$. The proof of the above theorem is established by showing that for any extreme point $\bb{v}$ of $\Ro{2}$, there exists an $\bb{r}\in \Ri{2}$ at distance less that $\sqrt{3}/2$ from $\bb{v}$.  Since any point $\bb{R}$ in $\Ro{2}$ is a convex combination of extreme points of $\Ro{2}$, we can employ a time-sharing protocol among the various achievable rate points $\{\bb{r}\}$ and achieve a rate point at distance less that $\sqrt{3}/2$ from $\bb{R}$.

\subsection{An approximate expression for the throughput.}

As an application of the above result, consider the symmetric scenario where $\Power_1 = \Power_2 = \Power$, and where each user is active with probability $p$. Based on the knowledge of $p$, transmitters optimize the choice of the encoding rates so that the throughput is maximized. Formally, we look for the solution of the following optimization problem:
\small
\begin{equation*}
\max  p(1-p)\left[ r_1(\{1\}) + r_2(\{2\}) \right] + p^2 \left[ r_1(\{1,2\}) + r_2(\{1,2\}) \right]
\end{equation*}
\normalsize
subject to the constraint that the rates should be in $\Ca{}$. Combining Theorem \ref{thm:out2f} and Theorem \ref{thm:in2f}, it is possible to show that the above maximum is equal $T(p,\Power) + \varepsilon(p,\Power)$, where
\begin{equation*}
T(p,\Power) = \left\{
  \begin{array}{ll}
    2 p (1-p)\C(\Power), & \hbox{if } p \in (0, \plo{2}{1}{\Power}]; \\
    p\C(2\Power), & \hbox{if } p \in (\plo{2}{1}{\Power},1],
  \end{array}
\right.
\end{equation*}
$\plo{2}{1}{\Power} = 1-\C(2\Power)/(2\C(\Power)) \in (0,1/2]$, and $0\le\varepsilon(p,\Power)\le 1$. Observe that the bound on the error term holds for any choice of the parameters $p$ and $\Power$.

The coding strategy used to achieve $T(p,\Power)$ is similar to the one described for the case of the symmetric BD channel. If the transmission probability $p$ lies in the interval $(0, \plo{2}{1}{\Power}]$, then user $i$ transmits message $W_{i,1}$ encoded at the maximum point-to-point coding rate, i.e., $\C(\Power)$. If, instead, the transmission probability $p$ lies in the interval $(\plo{2}{1}{\Power},1]$, then each active user transmits message $W_{i,2}$ encoded at rate $1/2\C(2\Power)$, i.e., at half the sum-rate capacity of the two-user AWGN-MAC. The parameter $\plo{2}{1}{\Power}$ represents a threshold value below which it is worth taking the risk of incurring in a packet collision. Observe that $\plo{2}{1}{\Power} \rightarrow 1/2$ as $\Power \rightarrow \infty$.
\begin{figure}
\begin{center}
\scalebox{1}{
\psfrag{p}{ {$p$}}
\psfrag{p1}{ {$\plo{2}{1}{\Power}$}}
\psfrag{A}{Adaptive rate approach \cite{elgamal}}
\psfrag{I}{Throughput (Inner bound)}
\psfrag{O}{Throughput (Outer bound)}
\psfrag{F}{Full CSI transmitters}
\psfrag{T}{bits/symbol}
\psfrag{C}{  $\C(2\Power)$}
\includegraphics[width=3.7in]{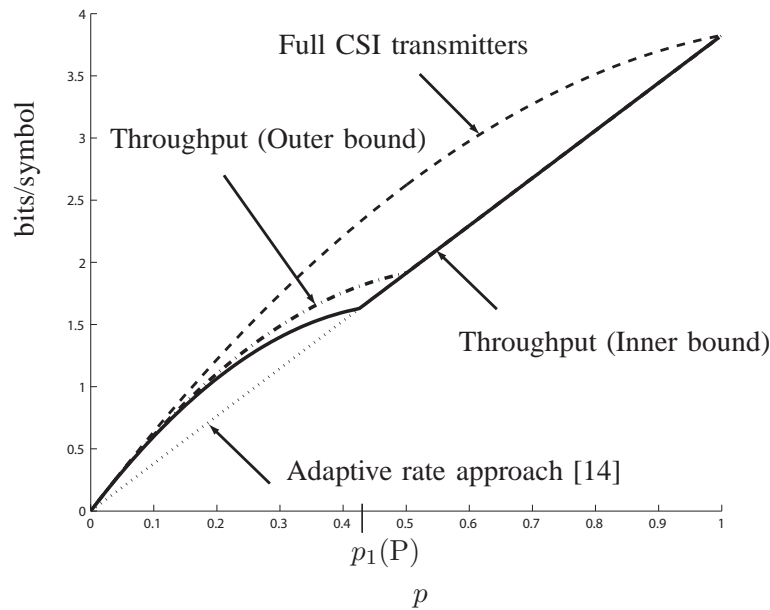}}
\end{center}
\caption{Throughput of the two-user symmetric AWGN-RAC ($\Power = 20 \text{dB}$).} \label{fig:sumrate}
\end{figure}

Fig.~\ref{fig:sumrate} compares $T(p,\Power)$ to the expected sum-rate achieved under the adaptive-rate framework~\cite{elgamal}, and to its counterpart assuming that full CSI is available to the transmitters. In the adaptive-rate framework, each sender transmits at a rate of $\C(2\Power)/2$, so that users can always be decoded. The figure illustrates how our approach allows us to improve upon the expected adaptive sum-rate for small values of $p$, for which the collision probability is small. In this regime,  our inner bound is in fact close to the curve obtained giving full CSI to the transmitters. Later in the paper, we shall see that the gain provided by our approach becomes more significant when the population size of the network increases.

\section{The $m$-user additive RAC}
\label{sec:m}
In this section we extend the analysis previously developed for a two-user system to the case of an $m$-user MAC, where $m$ denotes an integer $\ge 2$, and in which each transmitter can be in two modes of operation, active or not active.  The set of active users, denoted in the sequel by $A$, determines the \emph{state} of the channel. That is, the channel is said to be in state $A$ if all users in the set $A$ are active. As in the two-user case, transmitters only know their own state component, and encode data into independent streams of information. The common receiver knows the set of active users, and decodes subsets of the transmitted data streams depending on the state of the channel.

By introducing one auxiliary receiver per each channel state, we can map this problem to a broadcast network with $m$ transmitters and $2^m-1$ receivers. A one-to-one correspondence exists between the set of receivers and the set of non-empty subsets of $\{1,\dotsc,m\}$, so that for each set of active users $A$, there exists a unique corresponding receiver, which with abuse of notation we refer to as receiver $A$. Receiver $A$ observes the sum of the codewords transmitted by users in $A$ plus noise, and decodes a subset of the data streams sent by the active users. Observe that for a given channel state, only one among these auxiliary broadcast receivers corresponds to the actual physical receiver.

The formal description of the problem is as follows.

\subsection{Problem formulation}
\begin{defn}
An $m$-user DM-RAC $( \{\mathcal{X}_1,\dotsc,\mathcal{X}_m\},$ $\{\mathcal{Y}_A:A \varsubsetneq \{1,\cdots,m\} \},$ $(p(\{y_A: A \varsubsetneq \{1,\cdots,m\}\}|x_1,\dotsc,x_m) )$ consists of
$m$ input sets $\mathcal{X}_1,\dotsc,\mathcal{X}_m$, $2^m-1$ output sets $\{\mathcal{Y}_A\}$,  and a collection of conditional probabilities on the output sets.
\mbox{}
\end{defn}
The channel is \emph{additive} if at any discrete unit of time $t\in \{1,\dotsc,n\}$, the input symbols $(X_{1,t},\dotsc,X_{m,t})$ are
mapped into $2^{m}-1$ channel output symbols $\{Y_{A,t}\}$ via the additive map
\begin{equation}
\label{eq:prob}
Y_{A,t} = \sum_{a \in A } X_{a,t} + Z_{A,t},
\end{equation}
where the $\{Z_{A,t}\}$ are mutually independent random variables with values in a set $\mathcal{Z}$, and the sum is over a field $F$ such that there exists $m$ embeddings $F_i: \mathcal{X}_i \rightarrow F$, and one embedding $F_{m+1}: \mathcal{Z} \rightarrow F$.
In the next section we consider two classes of additive random access channels: the symmetric BD-RAC, for which the channel inputs are strings of bits, and the sum is binary; and the symmetric AWGN-RAC, for which $\mathcal{X}=\mathcal{Z}=\mathbb{R}$, the channel inputs are subject to an average power constraint, and the sum is over the reals.
\begin{defn}
A \emph{message structure} $\mathscr{W} = ( \{\mathcal{W}_1,\dotsc,\mathcal{W}_m\}, \{\mathcal{W}_i(A): i \in A \subseteq \{1,\cdots,m\}\})$ for an $m$-user RAC
consists of $m$ input message sets $\mathcal{W}_i$, $\mathcal{W}_i = \{ W_{i,1}, \cdots, W_{i,|\mathcal{W}_i|} \}$, and $m 2^{m-1}$ output sets $\mathcal{W}_i(A)$, $\mathcal{W}_i(A)\subseteq \mathcal{W}_i$, such that the following condition is satisfied:
\begin{enumerate}
  \item[\textbf{A1.}] $\mathcal{W}_i(B) \subseteq \mathcal{W}_1(A)$ for all $i \in  A \subseteq B \subseteq \{1,\dotsc,m\}$.
\end{enumerate}
\end{defn}
For each $i$ and $j \in \{1, \dotsc, |\mathcal{W}_i| \}$, message $W_{i,j}$ is a random variable independent of everything else and uniformly distributed over a set with cardinality $2^{n R_{i,j}}$, for some non-negative rate $R_{i,j}$, $j \in \{1, \dotsc, |\mathcal{W}_i| \}$.

The reason for imposing condition \textbf{A1.} is as follows. Observe from \eqref{eq:prob} that if $A \subseteq B$ and the marginal distributions of the noises $Z_B$ and $Z_A$ are equal, then $Y_B$ is a (stochastically) degraded version of $Y_A$. Then, condition \textbf{A1.} says that the ``better'' receiver $A$ \emph{must} decode what can be decoded at the ``worse'' receiver $B$.

For a given message structure $\mathscr{W}$, let
\begin{equation}
\label{eq:ria}
r_i(A)= \sum_{j:W_{i,j}\in \mathcal{W}_i(A)} R_{i,j}
\end{equation}
denote the sum of the rates of the messages in $\mathcal{W}_i(A)$. Observe that \eqref{eq:ria} defines a linear mapping from $\mathbb{R}_+^{|\mathcal{W}_1|\times \dotsc \times |\mathcal{W}_m|}$ into $\mathbb{R}_+^{m2^{m-1}}$ that shows how a \emph{macroscopic} quantity, the rate at which user $i$ communicates to receiver $A$, is related to various \emph{microscopic} quantities, the coding rates of the individual transmitted messages.

\begin{defn}
An \textit{$n$-code} for the RAC
$( \{\mathcal{X}_1,\dotsc,\mathcal{X}_m\},$ $\{\mathcal{Y}_A:A \varsubsetneq \{1,\cdots,m\} \},$ $(p(\{y_A: A \varsubsetneq \{1,\cdots,m\}\}|x_1,\dotsc,x_m) )$
and for the message structure $\mathscr{W}$
consists of $m$ encoding functions (encoders) and $2^m-1$ decoding functions (decoders). Encoder $i$ maps each $\{ W_{i,1}, \cdots, W_{i,|\mathcal{W}_i|} \}$ into a random codeword $\bb{X}_i \triangleq \{X_{i,1},X_{i,2},$ $\dotsc,X_{i,n}\}$ of $n$ random variables with values in the set $\mathcal{X}_i$. Decoder $A$ maps each channel output sequence $\bb{Y}_A \in \mathcal{Y}_A^n$ into a set of indexes $\cup_{j: W_{i,j} \in \mathcal{W}_{i}(A)} \{ \hat{W}_{i,j} \}$, where each index $\hat{W}_{i,j} \in \{1,\dotsc,2^{2nR_{i,j}}\}$ is an estimate of the corresponding transmitted message $W_{i,j} \in \mathcal{W}_{i}(A)$.
\mbox{}
\end{defn}

\begin{defn}
For a given $n$-code, the average \emph{probability of decoding error} at the decoder $A$ is defined as
\begin{align}
\label{eq:merror}
\text{Pr}\left \{ \hat{W}_{i,j} \neq W_{i,j} : W_{i,j} \in \mathcal{W}_{i}(A), j \in \{1,\dotsc, |\mathcal{W}_{i}(A)|\}, i \in A \right \}.
\end{align}
\mbox{}
\end{defn}
\begin{defn}
A rate tuple $\{ r_{i}(A) \}$ is said to be \emph{achievable} if there exists a sequence of $n$ codes such that
the average probability of a decoding error \eqref{eq:merror} for each decoder vanishes to zero as the block size $n$ tends to infinity.
\mbox{}
\end{defn}
\begin{defn}
The \emph{capacity region } $\Ca{\mathscr{W}}$ of the $m$-user RAC
$( \{\mathcal{X}_1,\dotsc,\mathcal{X}_m\},$ $\{\mathcal{Y}_A:A \varsubsetneq \{1,\cdots,m\} \},$ $(p(\{y_A: A \varsubsetneq \{1,\cdots,m\}\}|x_1,\dotsc,x_m) )$ for the message structure $\mathscr{W}$ is closure of the set of achievable rate vectors $\{ r_{i}(A) \} $.
\mbox{}
\end{defn}
Finally,
\begin{defn}
The capacity region $\Ca{}$ of the $m$-user  RAC
$( \{\mathcal{X}_1,\dotsc,\mathcal{X}_m\},$ $\{\mathcal{Y}_A:A \varsubsetneq \{1,\cdots,m\} \},$ $(p(\{y_A: A \varsubsetneq \{1,\cdots,m\}\}|x_1,\dotsc,x_m) )$
is defined as
$$\Ca{} = \text{closure} (\cup_{\mathscr{W}} \Ca{\mathscr{W}}).$$
\mbox{}
\end{defn}

\subsection{An outer bound to the capacity $\Ca{}$}

\begin{thm}
\label{thm:genout}
The capacity region $\Ca{}$ of the additive $m$-user additive RAC in \eqref{eq:prob} is contained inside the set of non-negative rate tuples satisfying
\begin{equation}
r_i(B)  \le r_i(A) \quad \text{for all } i \in B \subseteq A,
\label{eq:genout2a}
\end{equation}
and
\begin{equation}
\sum_{k=1}^K  \Ra{i_k}{ \{i_1 \dotsc i_k\} }  \leq I( X_{i_1},\dotsc, X_{i_K};Y_{ i_1 \dotsc i_K }),
\label{eq:genout2}
\end{equation}
for all $K \in \{1,\dotsc, m \}$ and $i_1\neq \dotso \neq i_m \in \{1,\dotsc,m\}$, and some joint distribution $p(q)p(x_1|q)\dotsm p(x_m|q)$, where $|Q|\leq m! \times m$.
\mbox{}
\end{thm}
\begin{proof}
See Appendix II.
\end{proof}

\emph{Remark 1:} In the special case of a network with two users, it is immediate to verify that the outer bound given by the above theorem reduces to the region given by Theorem \ref{thm:1} and Theorem \ref{thm:out2f} for the two-user BD-RAC and the two-user AWGN-RAC, respectively.

\emph{Remark 2:} An inspection of the proof of the above theorem shows that the additive channel model assumed in the theorem can be
replaced with a more general family of maps, namely with those channels with the property that, if $\bb{X}_{A'}$ is given, then $\bb{Y}_{A}$ is statistically equivalent to $\bb{Y}_{A\setminus A'}$, $A' \subseteq A$.

\emph{Remark 3:} Observe that \eqref{eq:genout2} gives $m$ constraints for any permutation of the set $\{1,\dotsc,m\}$, so it defines $m \times m !$ inequalities.

Equation \eqref{eq:genout2} can be obtained as follows.
Suppose that we fix a set of active users $i_1,\dotsc,i_K$, for some $K \in \{1,\dotsc,m\}$, and we provide the receiver observing $\bb{Y}_{ i_1 \dotsc i_K}$ with messages in the set $\cup_{r=1}^K \mathcal{W}_{i_{K-r+1}} \setminus \mathcal{W}_{i_{K-r+1}}(\{i_1\dotsc i_{K-r+1}\}) $ as side information. Suppose that this receiver decodes one user at the time, starting with user $i_K$ and progressing down to user $i_1$. Let us consider the first decoding step. By assumption, receiver $\{i_1 \dotsc i_K\}$ can decode information in $\mathcal{W}_{i_K}(\{i_1 \dotsc i_K\})$ so, given the side information $\mathcal{W}_{i_K} \setminus \mathcal{W}_{i_K}(\{i_1 \dotsc i_K\})$ it has full knowledge of $\mathcal{W}_{i_K}$, it can compute the codeword $\bb{X}_{i_K}$ transmitted by user $i_K$ and subtract it from the aggregate received signal, obtaining $\bb{Y}_{i_1\dotsc i_K} -\bb{X}_{i_K}=\bb{Y}_{i_1\dotsc i_{K-1} }$. Thus, at the end of the first decoding step the channel output observed by receiver $\{i_1 \dotsc i_{K}  \}$ is statistically equivalent to $\bb{Y}_{ i_1 \dotsc i_{K-1} }$. It follows that at the next decoding step it can decode information in $\mathcal{W}_{i_{K-1}}( \{i_1 \dotsc i_{K-1}\} )$.
By proceeding this way, at the $r$th iteration we obtain a sequence which is statistically equivalent to $\bb{Y}_{i_1 \dotsc i_{K-r+1} }$. Hence, receiver $\{\i_1 \dotsc i_{K}  \}$ can decode information in $\mathcal{W}_{i_{K-r+1}}(\{i_1 \dotsc i_{K-r+1}\})$, then make use of the side information $\mathcal{W}_{K-r+1} \setminus \mathcal{W}_{i_{K-r+1}}(\{i_1 \dotsc i_{K-r+1}\}  )$ to compute $\bb{X}_{i_{K-r+1}}$ and subtract it from the aggregate received signal before turning to decoding the next user. In other words, at the $r$th step of the iteration user $i_{K-r+1}$'s signal is only subject to interference from users $i_1,\dotsc,i_{k-r}$, as the signal of the remaining users has already been canceled. Therefore, user $i_{k-r+1}$ communicates to the receiver at a rate equal to $r_{i_{k-r+1}}(\{i_1\dotsc i_{k-r+1}\})$.

In summary, equation \eqref{eq:genout2} says that the sum of the communication rates across the $K$ iterations cannot exceed the mutual information between
the channel inputs on the transmitters side and the channel output on the receiver side, regardless of the permutation on the set of users originally chosen.


\subsection{The throughput of a RAC}

Assume that each user is active with probability $p$, independently of other users, and that $p$ is available to the encoders. In light of these assumptions,
\begin{defn}
The maximum expected sum-rate, or \emph{throughput}, of a RAC is defined as
\begin{equation}
\label{eq:T}
T(p,m) \triangleq \max  \sum_{A \subseteq \{1,\dotsc, m \}} p^{|A|}(1-p)^{m-|A|} \; \sum_{ i \in A} \Ra{i}{A}.
\end{equation}
\mbox{}
where the maximization is subject to the constraint that the rates should be in the capacity region $\Ca{}$ of that channel.
\end{defn}

The fact that each user is active with the same probability $p$ has one important consequence. By re-writing the objective function in \eqref{eq:T} as
\begin{equation*}
\sum_{k=1}^m p^k(1-p)^{m-k} \; \sum_{ \substack{A \subseteq \{1,\dotsc, m \} \\ |A|=k}} \sum_{ i \in A} \Ra{i}{A}
\end{equation*}
and defining
\begin{equation}
\rh{k} = \sum_{ \substack{A \subseteq \{1,\dotsc, m \} \\ |A|=k}} \sum_{ i \in A} \Ra{i}{A}, \quad k \in \{1,\dotsc,m\},
\label{eq:rho}
\end{equation}
it is clear that the objective function in \eqref{eq:T} depends only  on $\rh{1},\dotsc,\rh{m}$. It follows that in order to compute $T(p,m)$ it is sufficient to characterize the optimal tradeoff among these $m$ variables. This motivates the following definition
\begin{defn}
Let $\Ca{\rhb{}}$ denote the image of the capacity $\Ca{}$ of an $m$-user additive RAC under the linear transformation given by \eqref{eq:rho}.
\mbox{}
\end{defn}
It should be emphasized that the symmetry of the problem allow us to greatly reduce the complexity of the problem: instead of characterizing $\Ca{}$, which is a convex subset of $\mathbb{R}^{m 2^{m-1}}_+$,  it suffices to study the set $\Ca{\rhb{}}$, which is a convex subset of $\mathbb{R}^{m}_+$. Thus, we have that
\begin{equation}
\label{eq:T2}
T(p,m) =\max_{ \rh{1},\dotsc,\rh{m} \in \Ca{\rhb{}}}  \sum_{k=1}^m p^k(1-p)^{m-k} \rh{k}.
\end{equation}
In the sequel, outer and inner bounds on $\Ca{\rhb{}}$ are denoted by $\Ro{\rhb{}}$ and $\Ri{\rhb{}}$ respectively. In what follows, we denote by
\small
\begin{align*}
f_{m,k}(p) \triangleq \binom{m}{k} p^k (1-p)^{m-k}
\end{align*}
\normalsize
the probability of getting exactly $k$ successes in $m$ independent trials with success probability $p$, and we denote by
\small
\begin{align*}
F_{m,k}(p) \triangleq \sum_{i=0}^k f_{m,i}(p)
\end{align*}
\normalsize
the probability of getting at most $k$ successes.

\section{Example 1: the $m$-user symmetric BD-RAC}
\label{sec:mB}
In this section, we consider the $m$-user generalization of the symmetric BD-RAC considered in Section \ref{sec:2d}, where all transmitted codewords are shifted by the \emph{same} amount. This model represents an approximation of a wireless channel in which signals are received at the same power level.

Suppose the $\mathcal{X}$ and $\mathcal{Y}$ alphabets are each the set $\{0,1\}$, the additive channel \eqref{eq:prob} is noise-free, so $Z_A \equiv 0$, and the sum is over $\text{GF}(2)$. Observe that this is the $m$-user version of the channel model in \eqref{eq:2det} in the special case where $n_1=\dotsc=n_m=1$.
The codeword length is normalized to 1. As mentioned above, this channel model can be thought of as a natural generalization of the packet collision model widely used in the networking literature, where packets are always encoded at rate one, so that transmissions are successful only when there is one active user.
%
%
Theorem \ref{thm:genout} yields the following proposition.
\begin{prop}
\label{prop:det}
The capacity region $\Ca{}$ of the $m$-user symmetric BD-RAC is contained inside the set of $\{r_{i}(A)\}$ tuples satisfying
\begin{equation}
r_i(B)  \le r_i(A) \quad \text{for all } i \in  B \subseteq A,
\label{eq:det0}
\end{equation}
and
\begin{equation}
\sum_{k=1}^m  \Ra{i_k}{ \{i_1 \dotsc i_k\} }  \leq 1,
\label{eq:det1}
\end{equation}
for all $i_1\neq \dotso \neq i_m \in \{1,\dotsc,m\}$.
\mbox{}
\end{prop}

\subsection{The throughput of the symmetric BD-RAC}
Next, we turn to the problem of characterizing the throughput $T(p,m)$ for the symmetric BD-RAC. The following theorem provides the exact characterization of $\Ca{\rhb{}}$ for this channel.
\begin{thm}
\label{thm:genout2}
$\Ca{\rhb{}}$ for the $m$-user symmetric BD-RAC  is equal to the $(\rh{1},\dotsc,\rh{m})$ tuples satisfying
\begin{subequations}
\begin{align}
\label{eq:t1}
\frac{\rh{k}}{ k \mybinom{m}{k}{1} } \geq \frac{\rh{k+1}}{ (k+1) \mybinom{m}{k+1}{1} } \geq \dotso \geq \frac{\rh{m}}{ m \mybinom{m}{m}{1} } \geq  0,
\end{align}
and
\begin{align}
\label{eq:t2}
\sum_{k=1}^m  \frac{\rh{k}}{ k \mybinom{m}{k}{1} }   \leq 1.
\end{align}
\end{subequations}
\mbox{}
\end{thm}
\begin{proof}
See Appendix III.
\end{proof}

We outline the proof of the theorem as follows. The outer bound in the above theorem makes use of Proposition \ref{prop:det}. To prove the achievability, we show that $\Ca{\rhb{}}$ is equal to the image under the linear transformation given by \eqref{eq:rho} of the capacity region $\Ca{\mathscr{W}}$ of the $m$-user symmetric BD-RAC for the message structure $\mathscr{W}$ defined by
\begin{equation}
\label{eq:w1}
\mathcal{W}_i = \{W_{i,1},\dotsc,W_{i,m}\}, \quad i \in \{1,\dotsc,m\}
\end{equation}
and
\begin{equation}
\label{eq:w2}
\mathcal{W}_i(A) = \cup_{j \geq |A|} W_{i,j},
\end{equation}
for $i \in A \subseteq \{1,\dotsc,m\}$. This message structure is the natural generalization of the message structure used for the two-user BD-RAC. Each sender transmits $m$ independent messages, which are ordered according to the amount of interference which they can tolerate, so that message $W_{i,j}$ is decoded when there are less than $j$ interfering, regardless the identity of the interferers.

To prove the achievability of $\Ca{\rhb{}}$ using this message structure, we observe that $\Ca{\rhb{}}$ is the convex hull of $m$ extreme points, and that to achieve the $k$th  extreme points it suffices that user $i$ transmits a \emph{single} information message, namely $W_{i,k}$, encoded at rate $\tfrac{1}{k}$. Thus, a simple single-layer coding strategy can achieve all extreme points of $\Ca{\rhb{}}$, and the proof of the achievability is completed by means of a time-sharing argument.

Having an exact characterization of $\Ca{\rhb{}}$ at hands, we can  explicitly solve the throughput optimization problem. The main result of this section is given by the following theorem.
\begin{thm}
\label{thm:detthm1}
Let $\Pi_m$ represent the partition of the unit interval into the set of $m$ intervals
\begin{align*}
(p_{0},p_{1}],(p_{1},p_{2}],\dotsc,(p_{m-1},p_{m}],
\end{align*}
where $p_{0} \triangleq 0$, $p_{m} \triangleq 1$ and, for $0< k < m$, $p_{k}$ is defined as the unique solution in $\left(0,1\right)$ to the following polynomial equation in $p$
\begin{align}
\label{eq:eqd}
\frac{1}{k+1}F_{m-1,k}(p) = \frac{1}{k}F_{m-1,k-1}(p).
\end{align}
Then, the following facts hold
\begin{enumerate}
\item $p_{1}=\frac{1}{m}$, $p_{m-1}=\frac{1}{m^{1/(m-1)}}$, and $p_{} \in (0,\frac{k}{m})$ for $k \in \{2,\dotsc,m-2\}$.
\item The throughput of the $m$-user symmetric BD-RAC  is given by
\begin{align}
\label{eq:td}
T(p,m) =  \frac{mp}{k} F_{m-1,k}(p), \text{ if } p \in (p_{k-1}, p_{k}],
\end{align}
for $k \in \{1,\dotsc,m\}$.
\item $T(p,m)$ is achieved when all active senders transmit a \emph{single} message encoded at rate
\begin{align}
\label{eq:optd}
r(p) = \frac{1}{k}, \text{ if } p \in (p_{k-1}, p_{k}],
\end{align}
for $k \in \{1,\dotsc,m\}$.
\item $T(p,m)$ is a continuous function of $p$; it is concave and strictly increasing in each interval of the partition $\Pi_m$.
\end{enumerate}
\end{thm}
\begin{proof}
See Appendix IV.
\end{proof}

\emph{Remarks:} The above theorem says that $T(p,m)$ can be achieved by a coding strategy which does not require simultaneous transmission of multiple messages. Instead, each active user transmits a single message encoded at rate $r(p)$. Inspection of \eqref{eq:opt} reveals that $r(p)$ is a piecewise constant function of $p$, whose value depends on the transmission probability $p$. If $p$ is in the $k$th interval of the partition $\Pi_m$, then $r(p)$ is equal to $\tfrac{1}{k}$. Similarly, the corresponding achievable throughput $T(p,m)$ is a piecewise polynomial function of $p$. The boundary values of the partition, denoted by the sequence $\{p_k\}$, are given in semi-analytic form as solutions of \eqref{eq:eqd}, and closed form expressions are available only for some special values of $m$ and $k$. Nevertheless, Theorem \ref{thm:thm1} provides the upper bound $p_k< \frac{k}{m}$.

The structure of the solution is amenable to the following intuitive
interpretation. Based on the knowledge of $m$ and $p$, transmitters estimate the number of active users. More precisely, if $p$ is in the $k$th interval of the partition $\Pi_m$, i.e., $p_{k-1}<p \le p_k$, then transmitters estimate that there are $k$ active users. Since $p_k < \frac{k}{m}$, it is interesting to observe that the computed estimator is in general different from the maximum-likelihood estimator $\lfloor mp \rfloor$. Then, they encode their data at rate $\tfrac{1}{k}$, that is, each user requests an equal fraction of the $k$-user binary MAC sum-rate capacity. Clearly, there is a chance that the actual number of active users exceed $k$, in which case a collision occurs. Vice-versa, the scheme results in an inefficient use of the channel when the number of active users is less than $k$. However, this strategy represents the right balance between the risk of packet collisions and inefficiency.

It is interesting to note that when $p \le p_{k-1} $ the optimal strategy consists of encoding at rate $1$, i.e., at the maximum rate supported by the channel. As already remarked, this is the coding strategy used in the classic ALOHA protocol. Notice that since $p_{1}=\frac{1}{m}$, this strategy is optimal when the probability of being active is less that the inverse of the population size in the network. In this case, there is no advantage in exploiting the multi-user capability at the receiver. On the other hand, for $p > \frac{1}{m}$, the throughput of an ALOHA system is limited by packet collisions, which become more and more frequent as $p$ increases. In this regime, the encoding rate has to decrease in order to accommodate the presence, which become more and more likely as $p$ increases, of other potential active users.

\subsection{Throughput scaling for increasing values of $m$}
\label{sec:scaling}

If we let the population size $m$ grow while keeping $p$ constant, the law of large number implies that the number of active users concentrates around $mp$, so one would expect that the uncertainty about the number of active users decreases as $m$ increases. This intuition is confirmed by the following corollary, which states that the probability of collision tends to zero as $m$ grows to infinity.
\begin{cor}
\label{thm:thmscaling}
Let $p \in (0,1)$. Then, $\lim_{m \rightarrow \infty} T(p,m) = 1$.
\end{cor}

So far, we have been assuming that $p$ does not depend on $m$. Assume now that the total packet arrival rate in the system is $\lambda$, and let $p = \frac{\lambda}{m}$ be the arrival probability at each transmitting node. Let $T(\lambda)$ denote the throughput in the limit $m \rightarrow \infty$. Then, by applying the law of rare events to \eqref{eq:eqd} and \eqref{eq:td} we obtain the following corollary to Theorem \ref{thm:detthm1}.
\begin{cor}
\label{cor:detthm1}
Let $\lambda_{0} \triangleq 0$, $\lambda_{\infty} \triangleq \infty$ and, for $0< k < \infty$, let $\lambda_{k}$ be defined as the unique solution in $\left(0,\infty\right)$ to the following polynomial equation in $\lambda$
\begin{align*}
\frac{1}{k+1}\Gamma(k+1,\lambda) = \Gamma(k,\lambda)
\end{align*}
where $\Gamma(k+1,\lambda)$ is the incomplete gamma function. Then, as $m$ tends to infinity, the throughput is given by
\begin{align*}
T(\lambda) = \frac{\lambda}{k! k} \Gamma(k+1,\lambda) , \text{ if }  \lambda \in (\lambda_{k-1}, \lambda_{k}],
\end{align*}
for $k \in \mathbb{Z}$. The rate which attains the throughput is given by $r(\lambda) = \frac{1}{k}$, if $\lambda \in (\lambda_{k-1}, \lambda_{k}]$, $k \in \mathbb{Z}$.
Finally, $T(\lambda)$ is a continuous function of $\lambda$; it is concave and strictly increasing in each interval $(\lambda_{k-1}, \lambda_{k}]$, and $\lim_{\lambda \rightarrow \infty} T(\lambda) =1$.
\end{cor}

\begin{figure}
\begin{center}
\psfrag{p}[l]{$\lambda$}
\psfrag{e}[l]{$e^{-1}$}
\psfrag{I}[l]{$T(\lambda)$}
\psfrag{M}[l]{Slotted ALOHA $\lambda e^{-\lambda}$}
\psfrag{T}{bits/symbol}
\includegraphics[height=2.8in]{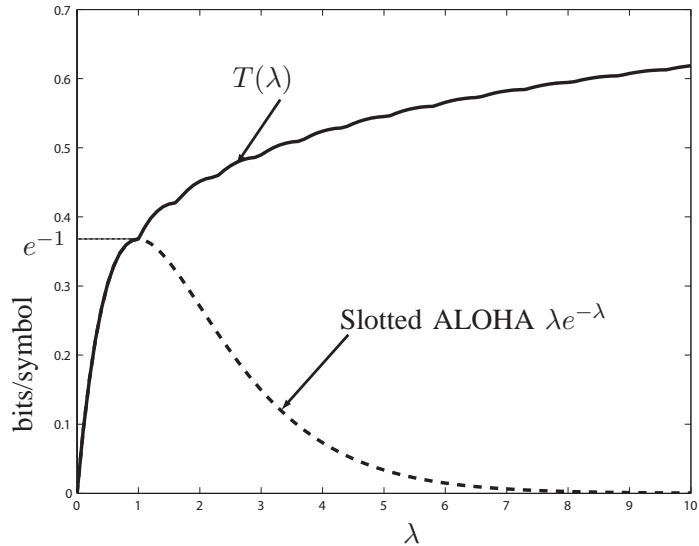}
\end{center}
\caption{Comparison between $T(\lambda)$ and the throughput of the slotted ALOHA protocol} \label{fig:2}
\end{figure}

Note that the claim above is in striking contrast with the throughput scaling of the classic slotted ALOHA protocol. The throughput of slotted ALOHA increases for small $\lambda$, it reaches a maximum $e^{-1}$ at $\lambda=1/m$, after which it
decreases to zero as $\lambda$ tends to infinity. See Fig.~\ref{fig:2} for a comparison between $T(\lambda)$ and the throughput of standard ALOHA as a function of $\lambda$.

\section{Example 2: the m-user symmetric AWGN-RAC}
\label{sec:mG}
We now turn to another important example of additive channels. Suppose that the codewords generated by the $m$ encoders are composed by $n$ random variables taking values over the reals, and whose realizations satisfy the following average power constraint
\begin{equation*}
\sum_{t=1}^n x_{i,t}^2 \leq n\Power
\end{equation*}
for some positive constant $\Power$. Observe that we focus on the \emph{symmetric} case in which all users are subject to the same received power constraint. Furthermore, suppose that $\{Z_A\}$ in \eqref{eq:prob} are independent standard Gaussian random variables, and that the sum in \eqref{eq:prob} is over the field of real numbers. Applying Theorem \ref{thm:genout}, we obtain the following proposition.
\begin{prop}
\label{prop:gau}
The capacity region $\Ca{}$ of the $m$-user symmetric AWGN-RAC is contained inside the set of $\{r_{i}(A)\}$ tuples satisfying
\begin{equation*}
r_i(B)  \le r_i(A) \quad \text{for all } i \in B \subseteq A,
\end{equation*}
and
\begin{equation*}
\sum_{k=1}^K  \Ra{i_k}{ \{i_1 \dotsc i_k\} }  \leq \C(K \Power),
\end{equation*}
for all $K \in \{1,\dotsc, m \}$ and $i_1\neq \dotso \neq i_m \in \{1,\dotsc,m\}$.
\mbox{}
\end{prop}

\subsection{An approximate expression to within one bit for the throughput}
Next, we turn to the problem of characterizing the throughput $T(p,m,\Power)$ for the symmetric AWGN-RAC as a function of the transmission probability $p$, the population size $m$, and the available power $\Power$. First, we provide inner and outer bounds on $\Ca{\rhb{}}$ for this channel.
\begin{thm}
\label{thm:gaurho}
Let $\Ro{\rhb{}}$ denote the set of rates
$\{\rh{k}\} \in \mathbb{R}^{m}$ such that
\begin{equation*}
\tfrac{\rh{k}}{ k \mybinom{m}{k}{1} } \geq \tfrac{\rh{k+1}}{ (k+1) \mybinom{m}{k+1}{1} } \geq \dotso \geq \tfrac{\rh{m}}{ m \mybinom{m}{m}{1} } \geq  0,
\end{equation*}
and
\begin{equation*}
\sum_{k=1}^K  \tfrac{\rh{k}}{ k \mybinom{m}{k}{1} }   \leq \C(K \Power),
\end{equation*}
for all $K \in\{1,\dotsc,m\}$. Let $\Ri{\rhb{}}$ denote the set of rates
$\{\rh{k}\} \in \mathbb{R}^{m}$ that satisfy \eqref{eq:t1} and
\begin{equation*}
\tfrac{1}{\C(\Power)}\tfrac{\rh{1}}{  \mybinom{m}{1}{1} }+ \sum_{k=2}^m  \left( \tfrac{k}{\C(k \Power)} - \tfrac{k-1}{\C((k-1) \Power)}\right)\tfrac{\rh{k}}{  k \mybinom{m}{k}{1} }   \leq 1.
\end{equation*}
Then,
$\Ri{\rhb{}} \subseteq \Ca{\rhb{}} \subseteq \Ro{\rhb{}}$.
\mbox{}
\end{thm}

The proof of the above theorem is omitted since it closely follows the proof of Theorem \ref{thm:genout2}. As for the case of the  BD-RAC, the achievable region in the above theorem is obtained by considering the message structure defined by \eqref{eq:w1} and \eqref{eq:w2} and the coding scheme we utilize does \emph{not} require the use of Gaussian superposition coding.

In virtue of Theorem \ref{thm:gaurho} it is possible to bound $T(p,m)$ as
\begin{equation*}
\label{eq:bb}
\underline{T}(p,m,\Power) \leq T(p,m) \leq \overline{T}(p,m,\Power),
\end{equation*}
where lower and upper bounds are given by \eqref{eq:T2} after replacing $\Ca{\rhb{},m}$ with $\Ri{\rhb{},m}$ and $\Ro{\rhb{},m}$ respectively.
The following theorem provides an expression for $\underline{T}(p,m,\Power)$.

\begin{thm}
\label{thm:thm1}
Let $\Pi_m(\Power)$ represent the partition of the unit interval into the set of $m$ intervals
\begin{align*}
(\plo{m}{0}{\Power},\plo{m}{1}{\Power}],\dotsc,(\plo{m}{m-1}{\Power},\plo{m}{m}{\Power}],
\end{align*}
where $\plo{m}{0}{\Power} \triangleq 0$, $\plo{m}{m}{\Power} \triangleq 1$ and, for $k \in \{1,\dotsc,m-1\}$, $\plo{m}{k}{\Power}$ is defined as the unique solution in $\left(0,\tfrac{k}{m} \right)$ to the following polynomial equation in $p$
\begin{align}
\label{eq:eq}
\frac{\C((k+1)\Power)}{k+1}F_{m-1,k}(p) = \frac{\C(k\Power)}{k}F_{m-1,k-1}(p).
\end{align}
Then, $\underline{T}(p,m,\Power)$ is a continuous function of $p$, concave, strictly increasing in each interval of the partition $\Pi_m(\Power)$, and is given by
\begin{align}
\underline{T}(p,m,\Power) = & \frac{\C(k\Power)}{k} mp  F_{m-1,k-1}(p) , \text{ if } p \in (\plo{m}{k-1}{\Power}, \plo{m}{k}{\Power}], \label{eq:TL}
\end{align}
for $k \in \{1,\dotsc,m\}$. To achieve $\underline{T}(p,m,\Power)$, it suffices that each active user transmits a unique message encoded at rate
\begin{align}
r(p,m,\Power) = & \frac{\C(k\Power)}{k} \text{ if } p \in (\plo{m}{k-1}{\Power}, \plo{m}{k}{\Power}], \label{eq:opt}
\end{align}
for $k \in \{1,\dotsc,m\}$.
\mbox{}
\end{thm}

The proof of the above theorem is omitted since it closely follows the proof of Theorem \ref{thm:genout2}. Similarly to what stated by Theorem \ref{thm:genout2} for the BD-RAC, the above theorem says that $\underline{T}(p,m,\Power)$ can be achieved by a coding strategy which does not require superposition coding: each active user transmits a single message encoded at rate $r(p,m,\Power)$. Both $r(p,m,\Power)$ and $\underline{T}(p,m,\Power)$ are piecewise constant function of $p$, whose value depends on the transmission probability $p$.

The coding scheme used to achieve $\underline{T}(p,m,\Power)$ for the symmetric AWGN-RAC is similar to the one used to achieve the throughput of the symmetric BD-RAC: based on the knowledge of $m$ and $\Power$ and $p$, transmitters estimate the number of active users. More precisely, if $p$ is in the $k$th interval of the partition $\Pi_m(\Power)$, i.e., $\plo{m}{k-1}{\Power}<p \le \plo{m}{k}{\Power}$, then transmitters estimate that there are $k$ active users.  Then, they encode their data at rate $\tfrac{1}{k}\C(k\Power)$, that is, each user requests an equal fraction of the $k$-user AWGN MAC sum-rate capacity.

A natural question to ask is how close this scheme is to the optimal performance. To answer this question, we first need to
provide an expression for $\overline{T}(p,m,\Power)$. This is done in the next Theorem.
\begin{thm}
\label{thm:thm1a}
Let $\Pi_m$ represent the partition of the unit interval into the set of $m$ intervals
\begin{align*}
(\pup{m}{0}{\Power},\pup{m}{1}{\Power}],\dotsc,(\pup{m}{m-1}{\Power},\pup{m}{m}{\Power}],
\end{align*}
where $\pup{m}{0}{\Power} \triangleq 0$, $\pup{m}{m}{\Power} \triangleq 1$ and, for every $k \in \{1,\dotsc,m\}$, $\pup{m}{k}{\Power}$ is defined as the unique solution in $\left(0,\tfrac{k}{m} \right)$ to the following polynomial equation in $p$
\begin{align}
\label{eq:equ}
\frac{1}{k+1}F_{m-1,k}(p) = \frac{1}{k}F_{m-1,k-1}(p).
\end{align}
Then, $\overline{T}(p,m,\Power)$ is a continuous function of $p$, concave and strictly increasing in each interval of the partition $\Pi_m(\Power)$, and is given by
\begin{align}
\overline{T}(p,m,\Power) = & mp \sum_{i =1}^m v_{k,i} F_{m-1,i-1}(p)
\text{ if } p \in (\pup{m}{k-1}{\Power}, \pup{m}{k}{\Power}], \label{eq:TLu}
\end{align}
for $k \in \{1,\dotsc,m\}$, where
\begin{equation}
\label{eq:ver1}
v_{1,i} = \left\{
\begin{array}{ll}
2\C(2\Power)-\C(\Power), & i =1, \\
2\C(i\Power)-\C((i+1)\Power)-2\C((i-1)\Power), & i \in \{2,\dotsc,m\}, \\
\C(m\Power)-\C((m-1)\Power), & i =m,
\end{array}
\right.
\end{equation}
For $k \in \{2,\dotsc,m-2\}$
\begin{equation}
\label{eq:verk}
v_{k,i} = \left\{
\begin{array}{ll}
0, & i \in \{1,\dotsc,k-1\}, \\
\frac{k+1}{k}\C(k\Power)-\C((k+1)\Power), & i =k, \\
2\C(i\Power)-\C((i+1)\Power)-\C((i-1)\Power), & i \in \{k+1,\dotsc,m-1\}, \\
\C(m\Power)-\C((m-1)\Power), & i =m,
\end{array}
\right.
\end{equation}
For $k =m-1$
\begin{equation}
\label{eq:verm1}
v_{m-1,i} = \left\{
\begin{array}{ll}
0, & i \in \{1,\dotsc,m-2\}, \\
\frac{m}{m-1}\C((m-1)\Power)-\C(m\Power), & i =m-1, \\
\C(m\Power)-\C((m-1)\Power), & i =m,
\end{array}
\right.
\end{equation}
For $k =m$
\begin{equation}
\label{eq:verm}
v_{m,i} = \left\{
\begin{array}{ll}
0, & i \in \{1,\dotsc,m-1\}, \\
\frac{1}{m}\C(m\Power), & i =m.
\end{array}
\right.
\end{equation}
\end{thm}
\begin{proof}
See Appendix V.
\end{proof}

The proof of the above theorem is conceptually simple but technical, as it requires finding the analytic solution of a linear program. Comparing the statements of Theorems \ref{thm:thm1} and \ref{thm:thm1a}, one can observe that the basic structure of $\overline{T}(p,m,\Power)$ and $\underline{T}(p,m,\Power)$ is the same. As opposed to the sequence $\{\plo{m}{k}{\Power}\}$ defined in Theorem \ref{thm:thm1}, the sequence $\{\pup{m}{k}{\Power}\}$ in Theorem \ref{thm:thm1a} does not depend on the power $\Power$. It is easy to see that $\plo{m}{k}{\Power}\le \pup{m}{k}{\Power} \le k/m$, for every $k$. Furthermore, the sequence $\{\pup{m}{k}{\Power}\}$ defined in Theorem \ref{thm:thm1a} is equal to the sequence defined in Theorem \ref{thm:genout2}. By directly comparing $\overline{T}(p,m,\Power)$ and $\underline{T}(p,m,\Power)$ we obtain the following result.
\begin{thm}
\label{thm:thm1b}
Let $p\in(0,1]$, $m\ge 2$ and $\Power>0$. Then,
\begin{align*}
\overline{T}(p,m,\Power) - \underline{T}(p,m,\Power)  \leq 1.
\end{align*}
\end{thm}
\begin{proof}
See Appendix VI.
\end{proof}

\begin{figure}
\begin{center}
\psfrag{p}[l]{$p$}
\psfrag{p31}[l]{$p_{3,1}$}
\psfrag{p32}[l]{$p_{3,2}$}
\psfrag{p41}[l]{$p_{1}(\Power)$}
\psfrag{p42}[l]{$p_{2}(\Power)$}
\psfrag{p43}[l]{$p_{3}(\Power)$}
\psfrag{rp}[l]{{\small $r(p,m,\Power)$}}
\psfrag{Tp}[l]{{\small $\overline{T}(p,m,\Power)$ vs $\underline{T}(p,m,\Power)$}}
\psfrag{Tu}[l]{{\small $\overline{T}(p,m,\Power)$}}
\psfrag{Tl}[l]{{\small $\underline{T}(p,m,\Power)$}}
\psfrag{tput}[l]{\hspace{-3em} {\small Throughput}}
\psfrag{rate}[l]{\hspace{-3em} {\small Rate allocation}}
\psfrag{C1}[l]{{\small $\C(\Power)$ }}
\psfrag{C2}[l]{{\small $\tfrac{1}{2}\C(2\Power)$ }}
\psfrag{C3}[l]{{\small $\tfrac{1}{3}\C(3\Power)$ }}
\psfrag{C4}[l]{{\small $\tfrac{1}{4}\C(4\Power)$ }}
\psfrag{D}[l]{{\tiny $\C(4\Power)$ }}
\psfrag{T}{bits/symbol}
\includegraphics[height=2.9in]{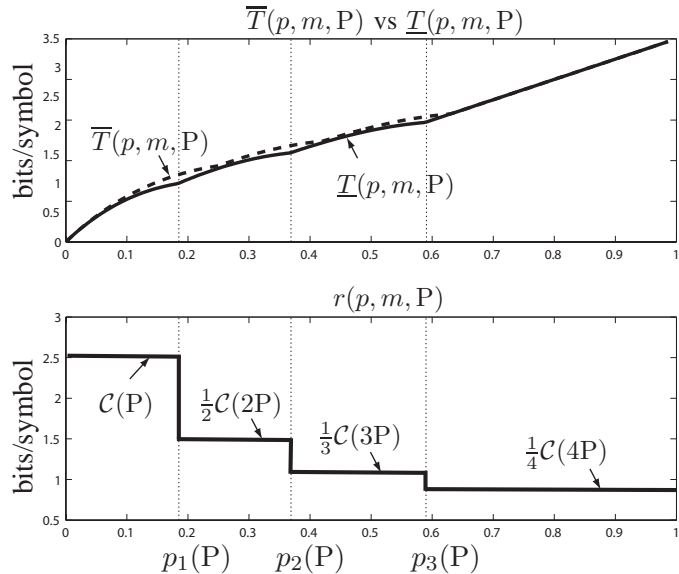}
\end{center}
\caption{Upper and lower bounds on the throughput of a four-user symmetric AWGN-RAC and encoding rate achieving the lower bound, as a function of the transmission probability $p$ ($\Power=15$ dB).} \label{fig:g}
\end{figure}
The above theorem says that our suggested coding scheme achieves an expected sum-rate which is only $1$ bit away from the optimum, independently of the values of $p$, $\Power$ and $m$. It it remarkable that the gap does not increase with the population size of the system.
Thus we conclude that transmitting at rate $\tfrac{1}{k}\C(k\Power)$ when $p$ is in the $k$th interval of the partition $\Pi_m(\Power)$ represents the right balance between risk of collision and efficiency: encoding rates above $\tfrac{1}{k}\C(k\Power)$ would increase the collision probability, yielding a decrease in the expected sum-rate. Viceversa, rates lower than $\tfrac{1}{k}\C(k\Power)$ would result in an inefficient use of the channel.

Fig.~\ref{fig:g} shows plots of $\overline{T}(p,m,\Power)$, $\underline{T}(p,m,\Power)$, and $r(p,m,\Power)$ for the case of networks with four users. Observe that the $\underline{T}(p,m,\Power)$ is a piecewise concave function of the transmission probability.

\subsection{Comparison with other notions of capacity}

The expression for the throughput derived in the previous section can be compared to similar expressions obtained assuming other notions of capacity. A natural outer bound is given by the throughput achieved assuming that full CSI is available to the transmitters. In this case, the sum-rate of the $k$-user AWGN-MAC can be achieved whenever $k$ users are active. Averaging over the message arrival probability, we obtain the following expression for the throughput:
\begin{align}
\label{eq:tcsi}
T_{CSI}(p,m,\Power) \triangleq \sum_{k=1}^m f_{m,k}(p)\C(k\Power).
\end{align}
On the other hand, if we study the symmetric AWGN-RAC following the adaptive capacity framework as in~\cite{elgamal}, then each transmitter designs a code which has to be decoded regardless the number of active users.
This is a conservative viewpoint and forces each user to choose a rate of $1/m \C(m \Power)$ so that users can be decoded even when all $m$ transmitters are active.
Thus, we obtain
\begin{align}
\label{eq:tad}
T_{AD}(p,m,\Power) \triangleq p \C(m\Power).
\end{align}
\begin{figure}
\begin{center}
\scalebox{1}{
\psfrag{p}{ {$p$}}
\psfrag{p1}{ {$\plo{2}{1}{\Power}$}}
\psfrag{A}{$T_{AD}(p,m,\Power)$}
\psfrag{I}{$\underline{T}(p,m,\Power)$}
\psfrag{O}{$\overline{T}(p,m,\Power)$}
\psfrag{F}{$T_{CSI}(p,m,\Power)$}
\psfrag{T}{bits/symbol}
\psfrag{C}{  $\C(2\Power)$}
\psfrag{M}{  MAP estimate for $|A|$}
\includegraphics[width=3.7in]{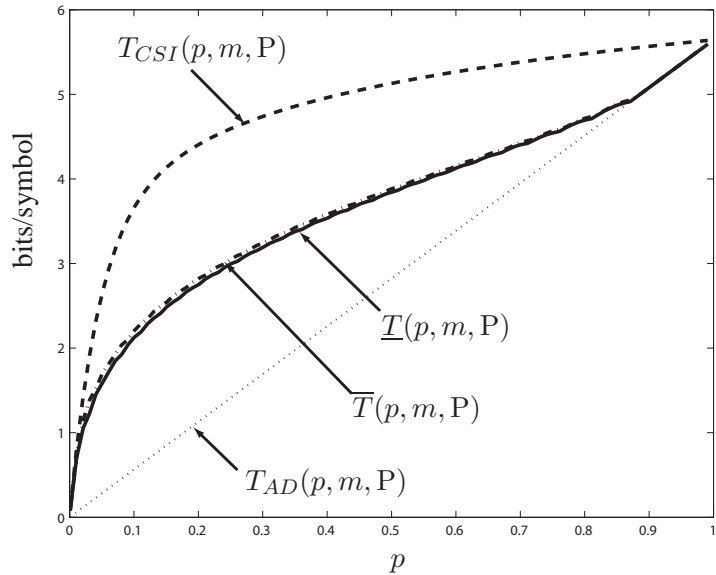}}
\end{center}
\caption{Throughput of the symmetric AWGN-RAC with $m=25$ users ($\Power = 20 \text{dB}$).} \label{fig:25user}
\end{figure}
Fig.~\ref{fig:25user} compares the obtained bounds on $T(p,m,\Power)$ for the case $m=25$ and $\Power=20\text{dB}$ to the
throughput under the adaptive-rate framework \eqref{eq:tad}, and assuming full CSI available to the transmitters \eqref{eq:tcsi}.

Finally, observe that in order to achieve $\underline{T}(p,m,\Power)$ transmitters have to estimate the number of active users by solving the polynomial equations \eqref{eq:eq}. A natural question to ask is what is the achievable throughput performance if a maximum-likelihood estimator for the number of active user is used instead. Consider the following strategy. Suppose that, based on the knowledge of $m$ and $p$, and assuming no prior on the number of active users, transmitters compute $k_{ML}$, the maximum-likelihood estimator for the number of active users, and encode their data at rate $\C(k_{ML}\Power)/k_{ML}$. Since the most probable outcome of $(m-1)$ Bernoulli trials\footnote{Each active transmitter estimates the state of the remaining $(m-1)$ users.} with success probability $p$ is the integer number between $mp-1$ and $mp$, we have that $k_{ML}=\lfloor m p \rfloor$. Thus, we obtain the following expression for the expected sum-rate capacity:
\begin{align}
\label{eq:tmap}
T_{ML}(p,m,\Power) \triangleq \frac{mp}{k_{ML}}\C(k_{ML}\Power) F_{m-1,k_{ML}-1}(p).
\end{align}
\begin{figure}
\begin{center}
\scalebox{.85}{
\psfrag{p}{ {$p$}}
\psfrag{p1}{ {$\plo{2}{1}{\Power}$}}
\psfrag{A}{$T_{AD}(p,m,\Power)$}
\psfrag{I}{$\underline{T}(p,m,\Power)$}
\psfrag{O}{$\overline{T}(p,m,\Power)$}
\psfrag{F}{$T_{CSI}(p,m,\Power)$}
\psfrag{T}{bits/symbol}
\psfrag{C}{  $\C(2\Power)$}
\psfrag{M}{ $T_{ML}(p,m,\Power)$ }
\includegraphics[width=3.7in]{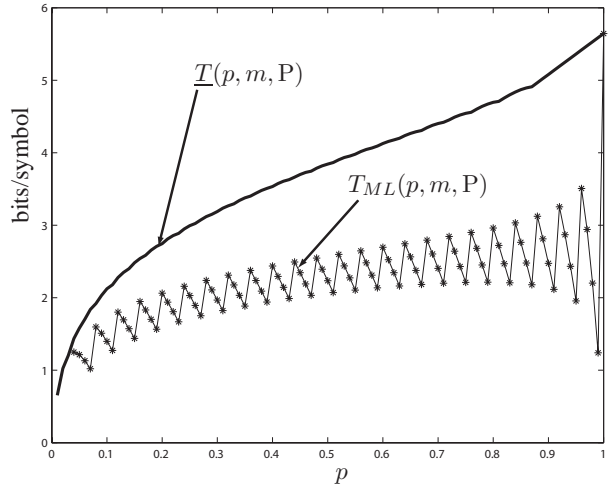}}
\end{center}
\caption{Throughput performance of the proposed estimator vs ML estimator for the number of active users ($\Power = 20 \text{dB}$, $m=25$).} \label{fig:25user_ML}
\end{figure}
Fig.~\ref{fig:25user_ML} compares $\underline{T}(p,m,\Power)$ and \eqref{eq:tmap} for the case $m=25$ and $\Power=20\text{dB}$. We remark is that the ML estimator for the number of active users result in a strictly suboptimal throughput performance.

\section{Discussion and practical considerations}

In networking, much research effort has been put in the design of distributed algorithms where each agent has limited information about the global state of the network. The model we developed in this paper allowed us to focus on the rate allocation problem that occurs when multiple nodes attempt to access a common medium, and when the set of active users is not available to the transmitters. Our analysis has lead to a distributed algorithm which is easily implementable in practical systems, and which is optimal in some information-theoretic sense. The rule of thumb which we have developed is that, upon transmission, senders should estimate the number of active users according to a prescribed algorithm based on the knowledge of the population size and the transmission probability $m$, and then choose the encoding rate accordingly.

In this paper we focused primarily on the problem of characterizing the throughput assuming perfect symmetry in the network, that is, the same transmission probability and received power constraint across users. The reasons for enforcing symmetry are twofold. First, throughput maximization is a meaningful performance metric only in symmetric scenarios. Second, it allows us to focus on random packet arrivals at the transmitters, and not on the different power levels at which transmitted signals are received by the common receiver. This set-up is a realistic model for uplink communications in power-controlled cellular wireless systems.
Nevertheless, an interesting open question is how to  apply the layering approach to the $m$-user AWGN-RAC with unequal power levels at the receiver, assuming that each sender only knows its own power level and state.

We made the underlying assumption that users can be synchronized, both at block and symbol level. In light of this assumption, a time-sharing protocol could be employed to prove achievability results. A simple way to achieve this partial form of cooperation among senders is to establish, prior to any transmission, that different coding schemes are used in different fractions of the transmission time. However, in practice achieving such complete synchronization may not be feasible. An interesting open question is to characterize the performance loss due to lack of synchronism. In this case, the resulting capacity region need not be convex, as for the collision model without feedback studied by Massey and Mathys \cite{Massey}.

We also assumed that the receiver has perfect CSI, that is, it knows the set of active users. The question, relevant in practice, of how the receiver can acquire such information is not discussed here, and we refer the reader to the recent studies of Fletcher \textit{et al.} \cite{Fletcher},  Angelosante \textit{et al.} \cite{Lops}, and Biglieri and Lops \cite{Biglieri2}, which address the issue using sparse signal representation techniques and random set theory.

Finally, in this paper the transmission probability $p$ and the number of users $m$ play a pivotal role in setting the encoding rate, and these quantities are supposed to be known at the transmitters. The probability $p$ is determined by the burstiness of the sources, while $m$ has to be communicated from the receiver to the transmitters. In practice, our model applies to communication scenarios in which the base station grants access to the uplink channel to $m$ users, but where only a subset of these users actually transmit data.

\section{Acknowledgment}
Prof. Young-Han Kim is gratefully acknowledged for many inspiring discussions during the course of the work reported here.

\section*{Appendix I: Proof of Theorem \ref{thm:2b} }

Observe that $\Ri{2}$ is a polytope in $\Reals^4_+$ defined as the intersection of eight hyperplanes, two of which representing
non-negativity constraints. By the Weyl-Minkowski theorem, $\Ri{2}$ is the convex hull of finitely many rate vectors. It is tedious but simple to verify that
\begin{align}
\Ri{2} & = \text{conv}
\left\{ \bb{v}_1, \dotsc,\bb{v}_{14} \right\} \nonumber \\
& = \text{conv}
\left\{
\begin{bmatrix}
 0 \\
 0 \\
 0 \\
 0
\end{bmatrix},\begin{bmatrix}
 \C(\Power_1) \\
 0 \\
 0 \\
 0
\end{bmatrix},\begin{bmatrix}
 0 \\
 \C(\Power_2) \\
 0 \\
 0
\end{bmatrix},\begin{bmatrix}
 \C(\Power_1) \\
 \C(\Power_2) \\
 0 \\
 0
\end{bmatrix},\begin{bmatrix}
 \C(\Power_1) \\
 0 \\
 \C(\Power_1) \\
 0
\end{bmatrix},\begin{bmatrix}
 0 \\
 \C(\Power_2) \\
 0 \\
 \C(\Power_2)
\end{bmatrix},\begin{bmatrix}
 \C\bigl(\tfrac{ \Power_1}{\Power_2+1}\bigr) \\
 \C(\Power_2) \\
 0 \\
 \C(\Power_2)
\end{bmatrix},\begin{bmatrix}
 \C(\Power_1) \\
 \C\bigl(\tfrac{\Power_2}{\Power_1+1}\bigr) \\
 0 \\
 \C\bigl(\tfrac{\Power_2}{\Power_1+1}\bigr)
\end{bmatrix},
\right. \nonumber \\
& \quad \quad \quad \left.
\begin{bmatrix}
 \C(\Power_1)  \\
 \C\bigl(\tfrac{\Power_2}{\Power_1+1}\bigr) \\
 \C(\Power_1) \\
 0
\end{bmatrix}, \begin{bmatrix}
 \C(\Power_1)  \\
 \C\bigl(\tfrac{\Power_2}{\Power_1+1}\bigr) \\
 \C(\Power_1) \\
 \C\bigl(\tfrac{\Power_2}{\Power_1+1}\bigr)
\end{bmatrix},\begin{bmatrix}
 \C\bigl(\tfrac{\Power_1}{\Power_2+1}\bigr) \\
 \C(\Power_2) \\
 \C\bigl(\tfrac{\Power_1}{\Power_2+1}\bigr) \\
 \C(\Power_2)
\end{bmatrix},\begin{bmatrix}
 \C(\Power_1)\\
 \C(\Power_2) \\
 \C\bigl(\tfrac{\Power_1}{\Power_2+1}\bigr) \\
 0
\end{bmatrix},\begin{bmatrix}
 \C(\Power_1) \\
 \C(\Power_2)\\
 0 \\
 \C\bigl(\tfrac{\Power_2}{\Power_1+1}\bigr)
\end{bmatrix},\begin{bmatrix}
 \C(\Power_1) \\
 \C(\Power_2)\\
 \C\bigl(\tfrac{\Power_1}{\Power_2+1}\bigr) \\
 \C\bigl(\tfrac{\Power_2}{\Power_1+1}\bigr)
\end{bmatrix}
\right\}. \label{eq:conv}
\end{align}
By convexity, it suffices to show that for every $i \in \{1,\dotsc,14\}$, there exists an achievable rate vector $\bb{r}_i$ such that $d(\bb{v}_i,\bb{r}_i) \le 1$. It is straightforward to verify that, for every $i \in \{1,\dotsc,11\}$, $\bb{v}_i \in \Ri{2}' \cup \Ri{2}''$. Thus, $d(\bb{v}_i,\bb{r}_i) = 0 $ for all $i \in \{1,\dotsc,11\}$.

Consider the rate vector $\bb{r}_{12} \in \Ri{2}'''$ obtained by setting equality sign in the inequalities \eqref{eq:in3} with $\beta_1 = \frac{\Power_2}{\Power_1} $ and $\beta_2 = 1$, i.e., $\bb{r}_{12} = \bigl[\C(\Power_2)+\C\left(\frac{\Power_1-\Power_2}{2\Power_2 +1}\right), \C(\Power_2), \C\left(\frac{\Power_1-\Power_2}{2\Power_2 +1}\right), 0\bigr]^\mathrm{T}$. We have that
\begin{align}
d(\bb{v}_{12},\bb{r}_{12}) & \leq  \sqrt{ \left| \C(\Power_1) - \C(\Power_2)- \C\left(\frac{\Power_1-\Power_2}{2\Power_2 +1}\right) \right|^2 + \left| \C\bigl(\tfrac{\Power_1}{\Power_2+1}\bigr) - \C\left(\frac{\Power_1-\Power_2}{2\Power_2 +1}\right) \right|^2
} \nonumber \\
& =  \sqrt{   \left| \frac{1}{2}  \log \left(1 + \frac{\Power_1 \Power_2 - \Power_2^2 }{\Power_1 \Power_2 - \Power_2^2}\right)  \right|^2 + \left| \frac{1}{2}  \log  \frac{2\Power_2+1}{\Power_2+1} \right|^2
} \nonumber \\
& \leq  \sqrt{   \left| \frac{1}{2}  \log \frac{2\Power_1\Power_2}{\Power_1 \Power_2 - \Power_2^2}  \right|^2 + \left| \frac{1}{2}  \log  \frac{2\Power_2+1}{\Power_2+1} \right|^2
} \nonumber \\
& \leq \frac{1}{\sqrt{2}}. \label{eq:d1}
\end{align}
Next, consider the rate vector $\bb{r}_{13} = [\C(\Power_1), \C(\Power_2),0, 0]^\mathrm{T} \in \Ri{2}'''$, obtained by setting equality sign in the inequalities \eqref{eq:in3} with $\beta_1 = 1 $ and $\beta_2 = 1$. We have that
\begin{align}
d(\bb{v}_{13},\bb{r}_{13}) = \left| \C\bigl(\tfrac{\Power_2}{\Power_1+1}\bigr) \right| \leq \frac{1}{\sqrt{2}}. \label{eq:d2}
\end{align}
Finally, the distance between $\bb{v}_{14}$ and $\bb{r}_{12}$ can be bounded as follows
\begin{align}
d(\bb{v}_{14},\bb{r}_{12}) & \leq  \sqrt{ \left| \C(\Power_1) - \C(\Power_2)- \C\left(\frac{\Power_1-\Power_2}{2\Power_2 +1}\right) \right|^2 + \left| \C\bigl(\tfrac{\Power_1}{\Power_2+1}\bigr) - \C\left(\frac{\Power_1-\Power_2}{2\Power_2 +1}\right) \right|^2 + \left| \C\bigl(\tfrac{\Power_2}{\Power_1+1}\bigr) \right|^2
} \nonumber \\
& \leq \frac{\sqrt{3}}{2}. \label{eq:d3}
\end{align}
Combining \eqref{eq:d1}, \eqref{eq:d2}, and \eqref{eq:d3} we conclude that $d(\bb{v}_{i},\bb{r}_{i})\leq \frac{\sqrt{3}}{2}$, $i \in \{ 12,13, 14\}$, which concludes the proof.

\section*{Appendix II: Proof of Theorem \ref{thm:genout}}
Inequalities \eqref{eq:genout2a} follow immediately from assumption \textbf{A1.}. Next, fix $i_1 \neq i_2 \neq \dotso \neq i_m \in \{1,\dotsc,m\}$. By Fano's inequality, we have that, for all $r \in \{1,\dotsc,m\}$,
\begin{align}
H\left( \mycup_{k=1}^r \D{i_k}{\{i_1 \dotsc  i_r\} }  \left| \bb{Y}_{ i_1 \dotsc i_{r} } \right. \right)  \leq n \epsilon_n,
\label{fano1}
\end{align}
where $\epsilon_n \rightarrow 0$ in the limit of $n$ going to infinity. In particular, \eqref{fano1} implies that
\begin{align}
H\left( \D{i_r}{\{i_1 \dotsc i_r\} }  \left| \bb{Y}_{ i_1 \dotsc i_{r} } \right. \right) \leq n \epsilon_n.
\label{fano2}
\end{align}
Let $K \in \{1,\dotsc,m\}$. Then, the following chain of equalities holds:
\begin{align}
& n\sum_{k=1}^K  \Ra{i_k}{ \{i_1 \dotsc i_k\} } \nonumber \\
%
%
& = H\left( \mycup_{k=1}^K \D{i_k}{\{i_1 \dotsc i_k\} } \right) \nonumber \\
%
& = H\left( \mycup_{k=1}^K \D{i_k}{\{i_1 \dotsc i_k\} } \left| \mycup_{k=1}^K \bigl\{ \mathcal{W}_{i_k} \setminus  \D{i_k}{\{i_1 \dotsc i_k\} } \bigr\}  \right. \right) \nonumber \\
%
%
& = I\left( \mycup_{k=1}^K \D{i_k}{\{i_1 \dotsc i_k\} }; \bb{Y}_{ i_1 \dotsc i_K } \left| \mycup_{k=1}^K \bigl\{ \mathcal{W}_{i_k} \setminus  \D{i_k}{\{i_1 \dotsc i_k\} } \bigr\}  \right. \right) \nonumber \\
& \; + H\left( \mycup_{k=1}^K \D{i_k}{\{i_1 \dotsc i_k\} } \left| \bb{Y}_{ i_1 \dotsc i_K }, \mycup_{k=1}^K \bigl\{ \mathcal{W}_{i_k} \setminus  \D{i_k}{\{i_1 \dotsc i_k\} } \bigr\} \right. \right) \label{eq:conv2}
\end{align}
The first term in the right hand side of \eqref{eq:conv2} can be upper bounded as follows
\begin{align}
& I\left( \mycup_{k=1}^K \D{i_k}{\{i_1 \dotsc i_k\} }; \bb{Y}_{ i_1 \dotsc i_K } \left| \mycup_{k=1}^K \bigl\{ \mathcal{W}_{i_k} \setminus  \D{i_k}{\{i_1 \dotsc i_k\} } \bigr\}  \right. \right) \nonumber \\
%
%
& = H\left( \bb{Y}_{ i_1 \dotsc i_K } \right) - H\left( \bb{Y}_{ i_1 \dotsc i_K } \left| \mycup_{k=1}^K \mathcal{W}_{i_k}  \right. \right) \nonumber \\
%
%
& \leq H\left( \bb{Y}_{ i_1 \dotsc i_K } \right) - H\left( \bb{Y}_{ i_1 \dotsc i_K } \left| \mycup_{k=1}^K \mathcal{W}_{i_k} ,\bb{X}_{i_1},\dotsc,\bb{X}_{i_{K}} \right. \right) \nonumber \\
%
%
%
& = \sum_{t=1}^n I\left( X_{i_1,t},\dotsc,X_{i_{K},t}; Y_ {i_1 \dotsc i_K,t} \right) \label{eq:conv2a}
\end{align}
where we use the fact conditioning reduces the entropy and the memoryless property of the channel. On the other hand, application of the chain rule on the second term at the left hand side of \eqref{eq:conv2} yields
\begin{align}
& H\left( \mycup_{k=1}^K \D{i_k}{\{i_1 \dotsc i_k\} } \left| \bb{Y}_{ i_1 \dotsc i_K }, \mycup_{k=1}^K \bigl\{ \mathcal{W}_{i_k} \setminus  \D{i_k}{\{i_1 \dotsc i_k\} } \bigr\} \right. \right) \nonumber \\
%
%
& = \sum_{r=1}^{K} H\left( \D{i_r}{\{i_1 \dotsc i_r\} } \left| \bb{Y}_{ i_1 \dotsc i_K }, \mycup_{k=1}^K \bigl\{ \mathcal{W}_{i_k} \setminus  \D{i_k}{\{i_1 \dotsc i_k\} } \bigr\} , \mycup_{k=r+1}^K \D{i_k}{\{i_1 \dotsc i_k\} } \right. \right) \nonumber \\
%
%
& = \sum_{r=1}^{K} H\left( \D{i_r}{\{i_1 \dotsc i_r\} } \left| \bb{Y}_{ i_1 \dotsc i_K }, \mycup_{k=1}^r \bigl\{ \mathcal{W}_{i_k} \setminus  \D{i_k}{\{i_1 \dotsc i_k\} } \bigr\} , \mycup_{k=r+1}^K  \mathcal{W}_{i_k} \right. \right) \nonumber \\
%
%
& = \sum_{r=1}^{K} H\left( \D{i_r}{\{i_1 \dotsc i_r\} } \left| \bb{Y}_{ i_1 \dotsc i_K }, \mycup_{k=1}^r \bigl\{ \mathcal{W}_{i_k} \setminus  \D{i_k}{\{i_1 \dotsc i_k\} } \bigr\} , \mycup_{k=r+1}^K  \mathcal{W}_{i_k} , \mycup_{k=r+1}^K \bb{X}_{i_k}  \right. \right) \label{eq:con3} \\
%
%
& = \sum_{r=1}^{K} H\left( \D{i_r}{\{i_1 \dotsc i_r\} } \left| \bb{Y}_{ i_1 \dotsc i_r }, \mycup_{k=1}^r \bigl\{ \mathcal{W}_{i_k} \setminus  \D{i_k}{\{i_1 \dotsc i_k\} } \bigr\} \right. \right) \nonumber \\
%
%
& = \sum_{r=1}^{K} H\left( \D{i_r}{\{i_1 \dotsc i_r\} } \left| \bb{Y}_{ i_1 \dotsc i_r } \right. \right) \label{eq:con4} \\
%
%
& \leq K n \epsilon_n ,\label{eq:conv5}
\end{align}
where \eqref{eq:con3} uses the fact that $\bb{X}_{i_k}$ is a function of $\mathcal{W}_{i_k}$, \eqref{eq:con4} uses the fact conditioning reduces the entropy, and \eqref{eq:conv5} follows from \eqref{fano2}.

Therefore, substituting \eqref{eq:conv2a} and \eqref{eq:conv5} into \eqref{eq:conv2}, we obtain that
\begin{align}
n\sum_{k=1}^K  \Ra{i_k}{ \{i_1 \dotsc i_k\} } \leq \sum_{t=1}^n I\left( X_{i_1,t},\dotsc,X_{i_{K},t}; Y_ {i_1 \dotsc i_K,t} \right) + n K \epsilon_n,
\end{align}
and the claim is completed by introducing a standard timesharing random variable and letting the block size $n$ tend to infinity.

\section*{Appendix III: Proof of Theorem \ref{thm:genout2} }

Let $\mathcal{P}$ denote the convex subset of $\mathbb{R}^m$ described described by inequalities \eqref{eq:t1} and \eqref{eq:t2}.
First we prove the converse part, by establishing that $\Ca{\rhb{}} \subseteq \mathcal{P}$. As a first step, we derive a useful identity. Let $k \in \{1,\dotsc,m\}$. Then,
\begin{align}
\sum_{ i_1\neq\cdots \neq i_m \in \{1,\dotsc, m \} } \Ra{i_k}{\{i_1 \dotsc i_k\}} & = (m-k)! \sum_{ i_1\neq\cdots \neq i_k \in \{1,\dotsc, m \} } \; \Ra{i_k}{\{i_1 \dotsc i_k\}} \nonumber \\
& = (m-k)! (k-1)! \sum_{ \substack{A \subseteq \{1,\dotsc, m \} \\ |A|=k}} \; \sum_{ i \in A} \Ra{i}{A} \nonumber \\
& = (m-k)! (k-1)! \rh{k}, \label{eq:th1}
\end{align}
where the second equality uses the fact that $\Ra{i_k}{i_1 \dotsc i_k} = \Ra{i_k}{\{i_{\sigma_1},\dotsc,i_{\sigma_{k-1}},i_k\}}$ for any permutation $\pmb{\sigma}$ over the set $\{1,\dotsc,k-1\}$. Now we can establish the necessity of \eqref{eq:t2}. It follows from \eqref{eq:det1} that the following inequality has to hold
\begin{align}
\label{eq:th0}
\sum_{k=1}^m  \Ra{i_k}{ i_1 \dotsc i_k }  \leq 1,
\end{align}
for all $i_1\neq \cdots \neq i_m \in \{1,\dotsc,m\}$. By summing both sides of $\eqref{eq:th0}$ over all permutations over the first $m$ integers, we obtain
\begin{align}
\sum_{ i_1\neq\cdots \neq i_m \in \{1,\dotsc, m \} } \sum_{k=1}^m  \Ra{i_k}{ \{i_1 \dotsc i_k\} } \leq m! \label{eq:th2}.
\end{align}
By means of \eqref{eq:th1}, \eqref{eq:th2} can be re-written as
\begin{align}
\sum_{k=1}^m (m-k)! (k-1)! \rh{k}  \leq m!  \label{eq:th3}.
\end{align}
Dividing both sides of \eqref{eq:th3} by $m!$, we conclude that \eqref{eq:t2} is a necessary condition for the achievability of a rate vector $\rhb{}$.

Next, note from \eqref{eq:det0} that $ \Ra{i}{A}  \geq \Ra{i}{B}$ for all $i \in A \subseteq B \subseteq \{1,\dotsc, m \}$ is a necessary condition to the achievability of a rate vector $\{ \Ra{i}{A} \}$. By summing these inequalities over all $B$ having cardinality $|A|+1$, we obtain that
\begin{align}
\Ra{i}{A}  \geq  \frac{1}{m-|A|} \sum_{ \substack{B: i \in A \subseteq B \subseteq \{1,\dotsc, m \} \\ |B|=|A|+1 }} \; \Ra{i}{B}. \label{eq:i0}
\end{align}
Next, observe that, for every $k \in \{1,\dotsc, m-1\}$,
\begin{align}
\rh{k}  & \geq  \sum_{ \substack{ A: A \subseteq \{1,\dotsc, m \} \\ |A|=k}} \; \sum_{ i \in A} \Ra{i}{A} \nonumber \\
& =  \sum_{i=1}^m \sum_{ \substack{A:  A \subseteq \{1,\dotsc, m \} \\ i \in A, \; |A|=k}} \; \Ra{i}{A} \nonumber  \\
& \geq \sum_{i=1}^m \sum_{ \substack{A:  A \subseteq \{1,\dotsc, m \} \\ i \in A, \; |A|=k}} \; \frac{1}{m-k} \sum_{ \substack{B: i \in A \subseteq B \subseteq \{1,\dotsc, m \} \\ |B|=k+1 }} \; \Ra{i}{B} \label{eq:i1} \\
& = \frac{1}{m-k} \sum_{i=1}^m \; \sum_{ \substack{B: B \subseteq \{1,\dotsc, m \} \\ i \in B, |B|=k+1 }} \; \sum_{ \substack{A:  A \subseteq B \\ i \in A, \; |A|=k}} \;   \; \Ra{i}{B} \label{eq:i2} \\
& = \frac{k}{m-k} \sum_{i=1}^m \; \sum_{ \substack{B: B \subseteq \{1,\dotsc, m \} \\ i \in B, |B|=k+1 }} \; \Ra{i}{B} \nonumber  \\
& = \frac{k}{m-k} \rh{k+1} \label{eq:i3}
\end{align}
where \eqref{eq:i1} follows from \eqref{eq:i0}, while \eqref{eq:i2} is obtained observing that there are $k$ subsets of $B$ which have  cardinality $k$ and contain the element $i$. After multiplying right and left hand side of \eqref{eq:i3} by $\frac{((m-1)!)}{(k-1)!}$ and rearranging the terms, we obtained the desired inequality
\begin{align*}
\frac{\rh{k}}{ k \mybinom{m}{k}{1} } \geq  \frac{\rh{k+1}}{ (k+1) \mybinom{m}{k+1}{1} }
\end{align*}
which proves \eqref{eq:t1}. In summary, we showed that inequalities \eqref{eq:t1} and \eqref{eq:t2} are necessary conditions for the achievability of a rate vector $\rhb{}$, i.e., $\Ca{\rhb{}} \subseteq \mathcal{P}$.

Next, we prove the achievability of $\mathcal{P}$, establishing the reversed inclusion $\mathcal{P} \subseteq \Ca{\rhb{}}$. To do so, it suffices to show that the extreme points of $\mathcal{P}$ are achievable, as the rest of the region can be achieved my means of a time-sharing protocol. We claim that
\begin{equation}
\mathcal{P} = \text{conv} \Bigl\{ \bb{0}, \Bigl\{ \frac{1}{k}\sum_{i=1}^k i \mybinom{m}{i}{1} \bb{e}_i \Bigr\}_{k=1}^m \Bigr\}. \label{eq:rirh2}
\end{equation}
where the vector $\bb{e}_i$ denotes the $i$th unit vector in $\mathbb{R}^{m}$.
To see this, consider an invertible linear transformation $L:\Reals^m\rightarrow \Reals^m$ given by
\begin{equation}
\left\{
\begin{array}{lll}
x_m & = \tfrac{\rh{m}}{ m \mybinom{m}{m}{1} }, & \\
x_k& = \tfrac{\rh{k}}{ k \mybinom{m}{k}{1} }-\tfrac{\rh{k+1}}{ (k+1) \mybinom{m}{k+1}{1} }, & k \in \{1,\dotsc,m-1\}.
\end{array}
\right.
\end{equation}
It is straightforward to check that the image $\mathcal{P}$ under $L$ is given by the oriented $m$-simplex $L \mathcal{P} = \{\bb{x}\in \Reals^m_+: \sum_{k=1}^m k x_k \le 1\} = \text{conv} \left\{ \bb{0},\bb{v}'_1,\dotsc, \bb{v}'_m \right\},$
wherein $\bb{v}'_k =\tfrac{1}{k} \bb{e}_k$.
Since $L$ is invertible, the extreme points of $\mathcal{P}$ can be obtained by applying $L^{-1}$ to the extreme points of $L\mathcal{P}$. Thus, $\mathcal{P} = \text{conv} \left\{ \bb{0}, \rhb{1},\dotsc, \rhb{m} \right\}$ where
\begin{equation}
\label{eq:v}
\rhb{k} =L^{-1} \bb{v}_k' = \frac{1}{k}\sum_{i=1}^k i \mybinom{m}{i}{0} \bb{e}_i,
\end{equation}
$k \in \{1,\dotsc,m\}$. Hence \eqref{eq:rirh2} is proved.

Next, we show that each rate vector $\rhb{k}$ given by \eqref{eq:v} is achievable. Consider the following message structure:
\begin{equation}
\label{eq:w1p}
\mathcal{W}_i = \{W_{i,1},\dotsc,W_{i,m}\}
\end{equation}
and
\begin{equation}
\label{eq:w2p}
\mathcal{W}_i(A) =
\left\{
  \begin{array}{ll}
    \cup_{j \geq |A|} W_{i,j}, & i \in A;\\
    \emptyset, & i \not \in A.
  \end{array}
\right.
\end{equation}
It is immediate to verify that the above sets satisfy conditions \textbf{A1.}, so the message structure is well defined.  For every $i$, sender $i$ transmits $m$ independent messages $\{W_{i,1},\dotsc,W_{i,m}\}$ encoded at rates $\{R_{i,1},\dotsc,R_{i,m}\}$. For every $k\in \{1,\dotsc,m\}$, the $k$th message $W_{i,k}$ is decoded at receiver $A$ if $i \in A$ and if $|A|\leq k$, that is, if user $i$ is active and there are less than $k$ active users.
To achieve the rate vector $\rhb{k}$ it suffices to set $R_{i,k}=\tfrac{1}{k}$ for all $i$, and the other rates equal to zero, that is, each sender $i$ transmits a \emph{single} message of information $W_{i,k}$ encoded at rate $\tfrac{1}{k}$. Encoding is performed by means of a standard multiple-access random codebook. It follows from \eqref{eq:w2p} that receiver $A$ decodes $W_{i,k}$ if $i \in A$ and $|A|\le k$. Thus, we have
\begin{equation}
\label{eq:w3}
r_i(A) =
\left\{
  \begin{array}{ll}
    \frac{1}{k}, & i \in A \text{ and } |A|\le k;\\
    0, & \text{otherwise}.
  \end{array}
\right.
\end{equation}
Observe that for every receiver $A$ the sum of the rates of the decoded messages is at most $1$. It follows that decoding can be performed by means of a standard $k$user multiple-access decoder. By plugging \eqref{eq:w3} into \eqref{eq:rho}, we obtain that
\begin{equation}
\rho_{k,i} =
\left\{
\begin{array}{lll}
\frac{i}{k} \mybinom{m}{i}{1}, & \text{if } i \in \{1,\dotsc, k\} \\
0, & \text{otherwise,}\\
\end{array}
\right.
\end{equation}
hence \eqref{eq:v} is achievable.

\section*{Appendix IV: Proof of Theorem \ref{thm:detthm1}}
In order to prove the theorem, we first need to state two lemmas. The first lemma builds upon properties of the cumulative distribution function of the Binomial distribution.
\begin{lem}
\label{lem1}
Let $k \in \{1,\dotsc,m-1\}$. There exists a $p_k \in \left(0,\frac{k}{m}\right)$ such that
\small
\begin{align}
\frac{1}{k}F_{m-1,k-1}(p) - \frac{1}{k+1}F_{m-1,k}(p) \; \left\{ \begin{array}{ll} > 0, & \quad  p<p_k \\ = 0, & \quad p=p_k \\ < 0, & \quad p>p_k \end{array}. \right. \label{eq:lem}
\end{align}
\normalsize
\end{lem}
\begin{IEEEproof}
Define $f(p) = \frac{1}{k}F_{m-1,k-1}(p) - \frac{1}{k+1}F_{m-1,k}(p)$. The binomial sum $F_{m-1,k-1}(p)$ is related to the incomplete Beta function by \cite[(6.6.4) page 263]{Abramowitz&Stegun:72}
\small
\begin{align}
\label{beta}
F_{m-1,k-1}(p) =  1 - k\mybinom{m-1}{k}{1} \int_{0}^p t^{k-1} (1-t)^{m-1-k}dt.
\end{align}
\normalsize
Substituting \eqref{beta} into the definition of $f(p)$ and differentiating, we obtain the following expression for the derivative of $f$ with respect to $p$ $f^\prime(p) =  -  \tfrac{1}{p(1-p)}f_{m-1,k}(p) \bigl[ 1 - p \tfrac{m}{k+1}\bigr]$.
By studying the sign of $f^\prime(p)$ one can see that $f(p)$ is a strictly decreasing function of $p$ in the range $\left(0,\tfrac{k+1}{m} \right)$, reaches a minimum at $p=\tfrac{k+1}{m}$ and is a strictly increasing in the interval $\left(\tfrac{k+1}{m},1\right)$. We have $f(1)=0$, and the Taylor expansion centered at $p=1$ shows that $f(p)$ increases to zero as $p$ tends to one. Thus, $f\left(\tfrac{k+1}{m}\right) < 0$. Note that $f(0) > 0$ so, by the monotonicity of $f$ and by the mean value theorem, there exists a unique $p_k \in \left(0,\tfrac{k+1}{m} \right)$ such that
\begin{align}
f(p)\; \left\{ \begin{array}{ll} > 0, & \quad  p<p_k \\ = 0, & \quad p=p_k \\ < 0, & \quad p>p_k \end{array}. \right. \label{eq:inlem}
\end{align}
To complete the proof, we show that $p_k<\frac{k}{m}$. Direct computation shows that $p_1 = 1/m$, while for $k \in \{2,\dotsc,m-1\}$, we have that
\begin{align}
f\bigl( \tfrac{k}{m} \bigr) & = \tfrac{1}{k} F_{m-1,k-1}\bigl( \tfrac{k}{m} \bigr) - \tfrac{1}{k+1} F_{m-1,k}\bigl( \tfrac{k}{m} \bigr) \nonumber \\
& \; = \left( \tfrac{1}{k} - \tfrac{1}{k+1} \right) F_{m-1,k-1}\bigl( \tfrac{k}{m} \bigr) - \tfrac{1}{k+1} f_{m-1,k}\bigl( \tfrac{k}{m} \bigr) \nonumber \\
& \; < \left( \tfrac{1}{k} - \tfrac{1}{k+1} \right) kf_{m-1,k-1}\bigl( \tfrac{k}{m} \bigr) - \tfrac{1}{k+1} f_{m-1,k-1}\bigl( \tfrac{k}{m} \bigr) \nonumber \\
& = 0, \label{eq:a}
\end{align}
where the inequality follows from the fact that $f_{m-1,i}\bigl( \tfrac{k}{m} \bigr)\leq f_{m-1,k-1}\bigl( \tfrac{k}{m} \bigr)$ for $i \in \{0,\dotsc,k-1\}$, with equality iff $i=k-1$, and that
$f_{m-1,k-1}\bigl( \tfrac{k}{m} \bigr)= f_{m-1,k}\bigl( \tfrac{k}{m} \bigr)$ for $k \in \{2,\dotsc,m-1\}$. Thus, \eqref{eq:inlem} and \eqref{eq:a} show that $p_k<\frac{k}{m}$ as claimed.
\end{IEEEproof}
Roughly speaking, the above says that to achieve the throughput the encoding rate has to decrease as the transmission probability increases. The second lemma shows that $1/k$ is the optimal encoding rate when $p$ is in the $k$th interval of the partition $\Pi_m(\Power)$.
\begin{lem}
\label{lem2}
Let  $k \in \{1,\dotsc,m\}$. Define $p_0\triangleq0$ and $p_m \triangleq 1$ and let $\{p_k\}_{k=1}^{m-1}$ be as in Lemma \ref{lem1}. Then,
\begin{align}
\frac{1}{k}F_{m-1,k-1}(p) \geq \frac{1}{j} F_{m-1,j-1}(p), \quad j \in \{1,\dotsc,m\},
\end{align}
for $p \in [p_{k-1}, p_k]$.
\end{lem}
\begin{IEEEproof}
In virtue of Lemma \ref{lem1}, it suffices to show that $p_k < p_{k+1}$, for $k \in \{0,\dotsc,m-1\}$. As $p_1 \in (0,1/m]$, it follows that $p_0<p_1$. Next, suppose that $k \in \{1,\dotsc,m-1\}$. Lemma \ref{lem1} shows that $\tfrac{1}{k} F_{m-1,k-1}(p_k)= \tfrac{1}{k+1} F_{m-1,k}(p_k)$ and that $p_k \in \left(0; \frac{k}{m}\right)$. Thus, we have
\begin{align}
& \tfrac{1}{k+1}F_{m-1,k}(p_k) - \tfrac{1}{k+2}F_{m-1,k+1}(p_k) \nonumber \\
& = \tfrac{1}{k}F_{m-1,k-1}(p_k) - \tfrac{1}{k+2}F_{m-1,k+1}(p_k) \nonumber \\
& = \left( \tfrac{1}{k} + \tfrac{1}{k+2}  \right) F_{m-1,k-1}(p_k) - \tfrac{1}{k+2}\bigl( F_{m-1,k-1}(p_k) + F_{m-1,k+1}(p_k)\bigr) \nonumber \\
& > \tfrac{2(k+1)}{k(k+2)} F_{m-1,k-1}(p_k) - \tfrac{2}{k+2}F_{m-1,k}(p_k) \nonumber \\
& = \tfrac{2(k+1)}{k(k+2)} F_{m-1,k-1}(p_k) - \tfrac{2(k+1)}{k(k+2)}F_{m-1,k-1}(p_k) \nonumber \\
& = 0, \label{eq:e}
\end{align}
where the inequality uses the fact that $F_{m-1,k-1}(p_k) + F_{m-1,k+1}(p_k)\bigr) < 2 F_{m-1,k}(p_k)$ for $p < k/m$. Comparing \eqref{eq:lem} and \eqref{eq:e}, we obtain the desired inequality $p_k < p_{k+1}$.
\end{IEEEproof}
Using the above lemma, it is immediate to prove theorem \ref{thm:detthm1}.
\begin{IEEEproof}
Observe that the optimum value of a linear program, if it exists, is always achieved at one of the extreme point of the feasibility set. Thus, \eqref{eq:rirh2} implies that
\begin{align*}
\underline{T}(p,m,\Power) & = \max_{k \in \{1,\dotsc,m\}} \frac{1}{k}\sum_{i=1}^k i \mybinom{m}{i}{0}  p^i (1-p)^{m-i} \\
& = \max_{k \in \{1,\dotsc,m\}} mp \frac{1}{k} F_{m-1,k-1}(p) \\
& = mp \sum_{k =1}^m\frac{1}{k} F_{m-1,k-1}(p) \mathbbm{1}_{\{  p \in (p_{k-1}, p_k] \}},
\end{align*}
where the last equality follows from Lemma \ref{lem2}.
\end{IEEEproof}
\section*{Appendix V: Proof of Theorem \ref{thm:thm1a}}

Let $c_k \triangleq \C(k\Power)$. In order to evaluate $\overline{T}(p,m,\Power)$, it is convenient to make the change of variable
\begin{equation}
\left\{
\begin{array}{lll}
x_m & = \tfrac{\rh{m}}{ m \mybinom{m}{m}{1} }, & \\
x_k& = \tfrac{\rh{k}}{ k \mybinom{m}{k}{1} }-\tfrac{\rh{k+1}}{ (k+1) \mybinom{m}{k+1}{1} }, & k \in \{1,\dotsc,m-1\}.
\end{array}
\right.
\end{equation}
Substituting the new variables into \eqref{eq:T2},  \eqref{eq:t1}, and \eqref{eq:t2} and performing a modicum of algebra, we obtain,
\begin{equation}
\label{eq:t4}
\overline{T}(p,m,\Power) = \max_{ \bb{x} \in \Ro{\bb{x},m} }  mp \sum_{i=1}^m F_{m-1,i-1}(p) \; x_i,
\end{equation}
where $\Ro{\bb{x},m}$ denote the set of rates $\{x_k\} \in \mathbb{R}^m_+$ such that
\begin{align}
\sum_{k=1}^{K-1}k  x_k + K  \sum_{k=K}^{m} x_k  \leq c_K \label{eq:t2b}
\end{align}
for every $K \in\{1,\dotsc,m\}$. Observe that the optimum value of the linear program \eqref{eq:t4} is achieved at one of the extreme point of the feasibility set. Therefore, to prove the theorem it suffices to show that $\{\bb{v}_k\}_{k=1}^m$ as defined in (\ref{eq:ver1}-\ref{eq:verm}) are extreme points of $\Ro{\bb{x},m}$, and that the objective function in \eqref{eq:t4} reaches a strict local maximum at $\bb{v}_k$ when $p $ is in the $k$th interval of the partition $\Pi_m(\Power)$.

For every $k \in \{1,\dotsc,m\}$, it is straightforward to check that $\bb{v}_k$ satisfies \eqref{eq:t2b} for $K \in\{k,\dotsc,m\}$, and that $\bb{v}_k$ has $k-1$ zero components. Thus, we conclude that $\bb{v}_k$ is an extreme point of $\Ro{\bb{x},m}$.

Next, we establish that if $p \in [\pup{m}{k-1}{\Power}, \pup{m}{k}{\Power}]$, where $\{\pup{m}{k-1}{\Power}\}$ are defined in Lemma \ref{lem1}, then the objective function reaches a local maximum at $\bb{v}_k$. We proceed by showing that the objective function at $\bb{v}_k$ is strictly greater than at any of its neighboring extreme points. By definition, two extreme points are neighbors if they are connected by an edge. It is possible to show that $\bb{v}_k$ has exactly $m$ neighbor extreme points, which we denote by $\left\{ \bb{n}_{j}^{(k)} \right\}_{j=1}^m$.
The proof of this fact is straightforward albeit fairly lengthy, so is not reported here. For $k \in\{1,\dotsc,m-1\}$, we have that
\begin{itemize}
 \item If $j \in \{1,\dotsc,k-1\}$, then
 \begin{equation}
 \label{eq:n1}
n_{j,i}^{(k)} = \left\{
\begin{array}{ll}
\frac{k}{j(k-j)}c_j - \frac{1}{k-j}c_k, & i = j, \\
0, & i \in \{1,\dotsc,j-1\} \cup \{j+1,\dotsc,k-1\}, \\
\frac{k-j+1}{j(k-j)}c_k - \frac{1}{k-j}c_j-c_{k+1}, & i = k, \\
v_{k,i}, & i \in \{k+1,\dotsc,m\}
\end{array}
\right.
\end{equation}
 \item If $j=k$, then $\bb{n}_{j}^{(k)} = \bb{v}_{k+1}$.
 \item If $j \in \{k+1,\dotsc,m-2\}$, then
   \label{eq:n2}
 \begin{equation}
n_{j,i}^{(k)} = \left\{
\begin{array}{ll}
\frac{3}{2}c_{j-1} - c_{j-2}- \frac{1}{2}c_{j+1} , & i = j-1, \\
0, & i=j, \\
\frac{3}{2}c_{j+1} - c_{j+2}- \frac{1}{2}c_{j-1} , & i = j+1, \\
v_{k,i}, & i \in \{1,\dotsc,j-2\} \cup \{j+2,\dotsc,m\}
\end{array}
\right.
\end{equation}
 \item If $j =m-1$, then
 \begin{equation}
   \label{eq:n3}
n_{m-1,i}^{(k)} = \left\{
\begin{array}{ll}
\frac{3}{2}c_{m-2} - c_{m-3}- \frac{1}{2}c_{m} , & i = m-2, \\
0, & i=m-1, \\
\frac{1}{2}c_{m}- \frac{1}{2}c_{m-2} , & i = m, \\
v_{k,i}, & i \in \{1,\dotsc,m-3\}
\end{array}
\right.
\end{equation}
\item Finally, if $j=m$ then
\begin{equation}
 \label{eq:n4}
n_{m,i}^{(k)} = \left\{
\begin{array}{ll}
c_{m-1} - c_{m-2}, & i = m-1, \\
0, & i =m, \\
v_{k,i}, & i \in \{1,\dotsc,m-2\},
\end{array}
\right.
\end{equation}
\end{itemize}
On the other hand, for $k=m$ and $j \in \{1,\dotsc, m-1\}$, we have that
 \begin{equation}
   \label{eq:n5}
n_{j,i}^{(m)} = \left\{
\begin{array}{ll}
\frac{m}{j(m-j)}c_{j} - \frac{1}{m-j}c_{m}, & i = j, \\
\frac{1}{m-j}c_{m} - \frac{1}{m-j}c_{j}, & i = m,
\end{array}
\right.
\end{equation}
It can be immediately verified that $\left\{ \bb{n}_{j}^{(k)} \right\}_{j=1}^m$ as defined above are extreme points of $\Ro{\bb{x},m}$, and neighbors of $\bb{v}_k$.

Next, we establish that the objective function in \eqref{eq:t4} reaches a local maximum at $\bb{v}_k$ by comparing the value achieved at $\bb{v}_k$ to the one at its neighboring extreme points. First, suppose $k \in \{1,\dotsc, m-1\}$.
\begin{itemize}
 \item  If $j \in \{1,\dotsc,k-1\}$, we can observe, from plugging \eqref{eq:n1} into \eqref{eq:t4} and performing some algebraic manipulations, that
\begin{align*}
& mp \sum_{i =1}^m (v_{k,i}-n_{j,i}^{(k)}) F_{m-1,i-1}(p) \\
& =   F_{m-1,k-1}(p) \left( \frac{1}{k-j}c_{j} - \frac{j}{k(k-j)} c_{k} \right) - F_{m-1,j-1}(p) \left( \frac{k}{j(k-j)}c_{j} - \frac{1}{k-j} c_{k} \right) \\
& >   F_{m-1,k-1}(p) \left( \frac{1}{k-j}c_{j} - \frac{j}{k(k-j)} c_{k} \right) - \frac{j}{k}F_{m-1,k-1}(p) \left( \frac{k}{j(k-j)}c_{j} - \frac{1}{k-j} c_{k} \right) \\
& = 0,
\end{align*}
because $F_{j-1} < \frac{j}{k} F_{m-1,k-1}(p) $ if $p $ is in the $k$th interval of the partition $\Pi_m(\Power)$.
 \item If $j=k$, then
\begin{align*}
& mp \sum_{i =1}^m (v_{k,i}-n_{k,i}^{(k)}) F_{m-1,i-1}(p) \\
&=   \frac{1}{k+1}\left( \frac{k+1}{k} c_{k} - c_{k-1} \right)  \left( \frac{1}{k}F_{m-1,k-1}(p) - \frac{1}{k+1}F_{m-1,k}(p) \right) \\
& > 0,
\end{align*}
 \item If $j \in \{k+1,\dotsc,m-1\}$, then
\begin{align*}
& mp \sum_{i =1}^m (v_{k,i}-n_{j,i}^{(k)}) F_{m-1,i-1}(p) =  \\
& = 2 \left(c_{j} - \frac{c_{j-1} + c_{j+1}}{2} \right) \left( F_{m-1,k}(p) - \frac{F_{m-1,k-1}(p) + F_{m-1,k+1}(p)}{2} \right) \\
& > 0,
\end{align*}
\item If $j=m$, then
\begin{align*}
mp \sum_{i =1}^m (v_{k,i}-n_{m,i}^{(k)}) F_{m-1,i-1}(p) & =   \left(c_{m} - c_{m-1} \right)  \left( F_{m-1,m-1}(p) - F_{m-1,m-2}(p) \right) \\
& > 0,
\end{align*}
\end{itemize}

Next, suppose $k =m$. Compare the utility function at $\bb{v}_m$ and $\bb{n}_j^{(m)}$.
\begin{align*}
& mp \sum_{i =1}^m (v_{k,i}-n_{m,i}^{(k)}) F_{m-1,i-1}(p) =   \\
& =   F_{m-1,m-1}(p) \left( \frac{1}{m-j}c_{j} - \frac{j}{m(m-j)} c_{m} \right) - F_{m-1,j-1}(p) \left( \frac{m}{j(m-j)}c_{j} - \frac{j}{m-j} c_{m} \right) \\
& =   F_{m-1,m-1}(p) \left( \frac{1}{m-j}c_{j} - \frac{j}{m(m-j)} c_{m} \right) - \frac{j}{m}F_{m-1,m-1}(p) \left( \frac{m}{j(m-j)}c_{j} - \frac{j}{m-j} c_{k} \right) \\
& = 0.
\end{align*}
Therefore, we have established that the objective function reaches a local maximum at $\bb{v}_k$ and completed the proof.

\section*{Appendix VI: Proof of Theorem \ref{thm:thm1b}}

Let $c_k \triangleq \C(k\Power)$. For every $k \in \{1,\dotsc,m\}$, if $p \in (\pup{m}{k-1}{\Power},\pup{m}{k}{\Power}]$ we have that
\begin{align}
\label{eq:f}
\underline{T}(p,m,\Power)  \ge mp \frac{c_k}{k}F_{m-1,k-1}(p).
\end{align}
In particular, equality holds in \eqref{eq:f} when $p \in [ \max(\pup{m}{k-1}{\Power}$ $,\plo{m}{k-1}{\Power}),\min(\pup{m}{k}{\Power},\plo{m}{k}{\Power}) ]$. It follows that
\begin{align}
\label{eq:a1}
\overline{T}(p,m,\Power) - \underline{T}(p,m,\Power)  \leq mp \sum_{i =1}^m v_{k,i} F_{m-1,i-1}(p)  - mp \frac{c_k}{k}F_{m-1,k-1}(p).
\end{align}
for $p \in (\pup{m}{k-1}{\Power},\pup{m}{k}{\Power}]$. To prove the theorem, we show that the right hand side of \eqref{eq:a1} is upper bounded by one for every $k \in \{1,\dotsc,m\}$. First, we consider the case $k=1$. By substituting \eqref{eq:ver1} into \eqref{eq:a1}, we obtain that
\begin{align*}
& mp \left[\sum_{i =1}^m v_{1,i} F_{m-1,i-1}(p)  - \C( \Power) F_{m-1,0}(p) \right] \\
& = mp \left[3 (c_2-c_1)F_{m-1,0}(p)  + \sum_{j=1}^m (c_{j+1}-c_{j})f_{m-1,j} \right] \\
& = mp \left[\frac{3}{2}F_{m-1,0}(p)  + \sum_{j=1}^{m-1} \frac{1}{2j}f_{m-1,j} \right] \\
& =  \frac{3}{2} f_{m,1}(p)  + \sum_{j=1}^{m-1} \frac{j+1}{2j}f_{m,j+1} \\
& \leq  \frac{3}{2} f_{m,1}(p)  + \sum_{j=2}^{m} f_{m,j} \\
& = \frac{3}{2} f_{m,1}(p) + \left( 1- f_{m,0}(p) - f_{m,1}(p) \right) \\
& \leq 1
\end{align*}
where the second equality uses the fact that $c_{j+1}-c_{j} \le 1/(2j)$,
while the last equality follows from $2f_{m,0}(p)\ge f_{m,1}(p)$ for $p \in (0,1/m]$. Similarly, from \eqref{eq:verk} we obtain that, for every $k\in \{2,\dotsc,m-1\}$,
\begin{align*}
& mp \left[\sum_{i =1}^m v_{k,i} F_{m-1,i-1}(p)  - \frac{c_k}{k}F_{m-1,k-1}(p) \right] \\
& = mp \left[ ( c_{k}- c_{k-1} ) F_{m-1,k-1}(p) +\sum_{i=k+1}^{m-1} ( 2c_{i}- c_{i-1} - c_{i+1})F_{m-1,i-1}(p) \right. \\
& \quad \quad  \left. + ( c_{m}- c_{m-1}) F_{m-1,m-1}(p)  \right] \nonumber \\
& = mp \left[ \left( c_{k}- c_{k-1}  + \sum_{i=k+1}^{m-1} \bigl( 2c_{i}- c_{i-1} - c_{i+1} \bigr) + c_{m}- c_{m-1}  \right) F_{m-1,k-1}(p) \right.  \\
& \quad \quad  \left. +\sum_{j=k+1}^{m} \sum_{i=j}^{m-1} \bigl( 2c_{i}- c_{i-1} - c_{i+1} + c_{m}- c_{m-1}  \bigr) f_{m-1,j-1}(p) \right] \nonumber \\
& = mp \sum_{j=k+1}^{m} (c_{j}-c_{j-1}) f_{m-1,j-1}(p) \\
& \le mp \sum_{j=k+1}^{m} \frac{1}{2(j-1)} f_{m-1,j-1}(p) \\
& = mp \sum_{j=k}^{m-1} \frac{1}{2j} f_{m-1,j}(p) \\
& \leq 1
\end{align*}
The proof is concluded observing that we have $\overline{T}(p,m,\Power) = \underline{T}(p,m,\Power) $ when $k=m$.

\bibliographystyle{ieeetr} 


\end{document}